\documentclass[paper,11pt]{JHEP}
\usepackage[centertags]{amsmath}
\usepackage{amsfonts}
\usepackage{amssymb}
\usepackage{amsthm}
\usepackage{graphicx}

\newcommand{\ket}[1]{|{#1}\rangle}

\newcommand{\ii}{\mathrm{i}}
\newcommand{\ee}{\mathrm{e}}
\newcommand{\one}{{\rm 1\kern -.9mm l}}                             %
\newcommand{\uno}{\mbox{1\!\negmedspace1}}

\newcommand{\Tr}{\mathrm{Tr}\,}

\title{Instantons in $\mathcal{N}=2$ magnetized D--brane worlds}
\author{Marco Bill\`o, Marialuisa Frau, Igor Pesando\\
Dipartimento di Fisica Teorica, Universit\`a di Torino\\
and I.N.F.N. - sezione di Torino \\
Via P. Giuria 1, I-10125 Torino, Italy
}
\author{Paolo Di Vecchia\\
The Niels Bohr Institute, Blegdamsvej 17, DK-2100 Copenhagen \O{},
Denmark\\
and NORDITA, Roslagstullsbacken 23, SE-10691 Stockholm, Sweden
}
\author{Alberto Lerda\\
Dipartimento di Scienze e Tecnologie Avanzate, Universit\`a del Piemonte Orientale\\
and I.N.F.N. - Gruppo Collegato di Alessandria - sezione di Torino\\
Via V. Bellini 25/G, I-15100 Alessandria, Italy
}
\author{Raffaele Marotta\\
I.N.F.N. - sezione di Napoli
and Dipartimento di Scienze Fisiche, Universit\`a di Napoli.\\
Complesso Universitario Monte S. Angelo - ed. G\\
Via Cintia - I-80126 Napoli, Italy}

\abstract{In a toroidal orbifold of type IIB string
theory we study instanton effects in $\mathcal{N}=2$ super Yang-Mills theories
engineered with systems of wrapped magnetized D9 branes and Euclidean D5
branes. We analyze the various open string sectors in this brane system and study
the 1-loop amplitudes described by annulus diagrams
with mixed boundary conditions, explaining their r\^ole in the stringy
instanton calculus. We show in particular that the non-holomorphic terms in these annulus
amplitudes precisely
reconstruct the appropriate K\"ahler metric factors that are
needed to write the instanton correlators in terms of
purely holomorphic variables. We also
explicitly derive the correct holomorphic structure of the
instanton induced low energy effective action in the Coulomb branch.
}
\preprint{DFTT/12/2007\\
NORDITA-2007-24}

\begin{document}

\section{Introduction}
\label{sec:intro}
In their original formulation  string theories were defined
as  perturbative expansions in the  string coupling constant $g_s$
that reproduce the corresponding perturbative field theoretical
expressions in the zero-slope limit ($\alpha' \rightarrow 0$).
For a long time it seemed very difficult, or even impossible, to reproduce in string theory
the non-perturbative effects that were instead known from field theory, such as
for example instanton effects.

A first important step in this direction was performed in
Ref. \cite{Polchinski:1994fq}, but it was only after the discovery of
string dualities and M theory that a real progress could be achieved.
In fact, by exploiting string dualities
it became clear that perturbative phenomena in one theory often correspond
to non-perturbative ones in the dual theory and {vice-versa},
and that the dependence on the string coupling constant of these
non-perturbative
effects is of the same type produced by instantons in field theory \cite{Becker:1995kb}.
Non-perturbative phenomena of this kind were discovered both in
type II theories~\cite{Green:1997di,Green:1997as,Kiritsis:1997em} and in the framework of
Heterotic/Type I  duality~\cite{Bachas:1997mc}.

These developments opened the way to a more systematic analysis
of instanton effects in string theory~\cite{Green:1997tv,Green:2000ke}.
Among the stringy non-perturbative configurations, the so-called
D-instantons, {\it i.e.} the D(--1) branes of type IIB, were the mostly studied ones
at the beginning and, after the discovery of the AdS/CFT correspondence,
they were intensively used to get additional evidence of the
equivalence between ${\mathcal{N}}=4$ super Yang-Mills theory (SYM)
in four dimensions and type IIB string theory on $AdS_5
\times S^5$~\cite{Banks:1998nr}--\cite{Dorey:1998xe}.

These results were largely based on the fact that
the instanton sectors of
${\mathcal N}=4$ SYM theory
can be described in string theory by systems of
D3 and D(--1) branes (or D-instantons)~\cite{Witten:1995im,Douglas:1996uz,Green:2000ke}.
In fact, the excitations of the
open strings stretching between two D(--1) branes,
or between a D3 brane and a D-instanton, are in one-to-one
correspondence with the moduli of the SYM instantons in the so-called
ADHM construction (for comprehensive reviews on the subject see, for example,
Refs.~\cite{Dorey:2002ik,Bianchi:2007ft}).
This observation can be further substantiated~\cite{Billo:2002hm} by showing
that the tree-level string scattering amplitudes on disks
with mixed boundary conditions for a D3/D(--1) system lead, in the
$\alpha^\prime\to 0$ limit, to the effective action on the instanton moduli
space of the SYM theory. Moreover, it can be proved~\cite{Billo:2002hm}
that the same disk diagrams also yield the classical profile
of the gauge instanton solution, in close analogy with the procedure
that generates the profile of the classical supergravity
D brane solutions from boundary states~\cite{Di Vecchia:1997pr}.

This approach can be easily adapted to
describe gauge instantons in SYM theories
with reduced supersymmetry by placing the D3/D(--1) systems
at suitable orbifold singularities.
It is also possible to take into account the
deformations induced by non-trivial gravitational
backgrounds both of NS-NS and  R-R type \cite{Billo:2004zq,Billo:2005fg,Billo:2006jm}.
For instance, by studying a D3/D(--1) system in an $\mathcal{N}=2$
orbifold and in the presence of a graviphoton background
it is possible to systematically obtain the instanton induced
gravitational corrections to the $\mathcal{N}=2$ low-energy
effective SYM action using perturbative string methods
\cite{Billo:2006jm}.

More recently, the string description of instantons has lead to new developments
that have received a lot of attention. In fact it has been shown in several
different contexts ~\cite{Beasley:2005iu}--\cite{Blumenhagen:2007bn} that the
stringy instantons may dynamically generate new types of superpotential terms in
the low-energy effective action of the SYM theory. These new types of F-terms
may have very interesting phenomenological implications, most notably they can
provide a mechanism for generating Majorana masses for
neutrinos~\cite{Blumenhagen:2006xt,Ibanez:2006da} in some semi-realistic string
extensions of the Standard Model.

However, one of the problems that one has to face in this approach
is that a superpotential term must be holomorphic in the
appropriate field theory variables, but what is holomorphic
in string theory is not quite the same of what is holomorphic
in supergravity. If we limit ourselves to a toroidal
compactification of string theory of the type
$\mathbb{R}^{1,3}\times\mathcal{T}_2^{(1)}
\times\mathcal{T}_2^{(2)}\times\mathcal{T}_2^{(3)}$,
the holomorphic quantities that naturally appear are
the complex  structures and the K\"ahler structures of the three tori,
together with the ten-dimensional axion-dilaton field. On the other hand,
when we incorporate the results of the string compactification
in a four-dimensional supergravity Lagrangian, the appropriate
fields to be used are different from those mentioned above and are
obtained from these by forming specific combinations with various
R-R fields (see, for instance, Ref.~\cite{Blumenhagen:2006ci}
for a review). Only when written in terms of these supergravity variables, the
F-terms have the correct holomorphic structure and the tree-level
SYM coupling constant is the sum of a holomorphic and an anti-holomorphic
quantity as required by supersymmetry.
When 1-loop effects are included, some non-holomorphic terms appear due
to the presence of massless modes which require an IR
regularization procedure, but they turn out to precisely
reconstruct the K\"ahler metrics of the various low-energy fields
~\cite{Kaplunovsky:1994fg,Louis:1996ya}~\footnote{See also Refs.
\cite{Shifman:1986zi}--\cite{de Wit:1995zg}.},
so that they can be re-absorbed with field redefinitions.

A similar pattern should occur also for the non-perturbative
F-terms induced by instantons in string models. While the holomorphic dependence
of these instanton contributions from the complex quantities of the
low-energy theory is a consequence of
the cohomology properties of the integration measure on the instanton
moduli space \cite{Hollowood:2002ds,Dorey:2002ik,Billo:2006jm},
the holomorphic dependence on the compactification moduli is not
at all obvious. This problem has
started to be analyzed only recently in the framework of
intersecting brane worlds in type IIA string
theory~\cite{Akerblom:2007uc}.

In this paper we consider instead a
toroidal orbifold compactification of type IIB string theory
in $\mathbb{R}^{1,3}\times\frac{\mathcal{T}_2^{(1)}
\times\mathcal{T}_2^{(2)}}{\mathbb{Z}_2}\times\mathcal{T}_2^{(3)}$
and study systems of fractional D9 branes that are wrapped and
magnetized on the three tori in such a way to engineer a
${\cal{N}}=2$ SYM theory with $N_F$ flavors. In particular we
distinguish the color D9 branes, which support the
degrees of freedom of the gauge multiplet, and the flavor D9
branes, which instead give rise to hyper-multiplets in the fundamental
representation. To study instanton effects in this set-up, we add a
stack of Euclidean D5 branes (E5 branes for short)
that completely wrap the internal manifold and hence describe point-like
configurations from the four-dimensional point of view~\footnote{These D9/E5
systems are essentially a T-dual version of the D3/D(--1) systems mentioned
above.}. If the
wrapping numbers and magnetization of these E5 branes are the same
as those of the color D9 branes, we have a stringy
realization of ordinary gauge theory instantons; if instead the internal
structure
of the wrapped E5 branes differs from that of the color branes,
then we have exotic instanton configurations of truly stringy nature. In
this paper we will consider the first case, but in principle our
results can be useful also to study the exotic cases.
The physical excitations corresponding to open strings
with at least one end-point on the E5 branes describe the
instanton moduli, and their mutual interactions, as well as their
couplings with the gauge and matter fields, can be explicitly
obtained from the $\alpha'\to 0$ limit of disk diagrams with mixed
boundary conditions, in complete analogy with the $\mathcal{N}=2$
system studied in Ref.~\cite{Billo:2006jm} in a non-compact orbifold.
In our case, however, we have to take into account also
the contribution of the compact internal space, and in particular
of its complex and K\"ahler structure moduli which explicitly
appear in the 1-loop amplitudes corresponding to annulus diagrams
with one boundary on the instantonic E5 branes and the other on the
D9 branes. We show with very general arguments that in supersymmetric
gauge theories these annulus diagrams with mixed boundary conditions describe
precisely
the 1-loop correction to the gauge coupling constant, in agreement
with some recent observations \cite{Abel:2006yk,Akerblom:2006hx}.
Besides the usual logarithmic terms that are responsible for the running of the
coupling constant, these 1-loop corrections in general contain
also some finite terms that are interpreted as threshold effects
\cite{Kaplunovsky:1994fg,Lust:2003ky}.
While in the non-compact orbifolds these thresholds are absent
\cite{Douglas:1996du,Bachas:1996zt,Di Vecchia:2005vm}, in the non-compact case
they give, instead, a relevant contribution and actually
produce crucial non-holomorphic terms that
precisely reconstruct the appropriate K\"ahler metric factors
which compensate those arising in the transformation from the
string to the supergravity basis. In this way one can explicitly prove that
the instanton induced low-energy effective action, when written in the
supergravity variables, has the correct holomorphic properties, as required
by supersymmetry.

The paper is organized as follows. In Section \ref{sec:n2_mod} we
review how to engineer $\mathcal{N}=2$ SYM theories with flavors using wrapped
magnetized D9 branes in a toroidal orbifold compactification of
type II string theory and discuss the relation between the string
basis and the supergravity basis which allows to determine the form of
the K\"ahler metric for the various scalar fields of the model.
In Section \ref{sec:inst_calc} we describe the instanton calculus
in string theory and discuss how to obtain the instanton induced
contributions to the low-energy effective action from
disk amplitudes. We also show how the 1-loop annulus amplitudes enter
in the calculation. Section \ref{sec:mix_ann} is devoted to perform the explicit
computation of these annulus amplitudes and to explain their r\^ole in the
instanton
calculus. In Section \ref{sec:hol_life} we show that the
non-perturbative effective actions generated by the E5 branes
have the correct holomorphic structure required by supersymmetry
for Wilsonian actions, if the appropriate variables of the
supergravity basis are used. Finally in Section \ref{sec:concl} we present our
conclusions and
in the Appendix we provide some technical details for the integral
appearing in the annulus amplitudes.

\section{$\mathcal{N}=2$ models from magnetized branes}
\label{sec:n2_mod}

In this section we review how to obtain gauge theories with $\mathcal{N}=2$
supersymmetry from systems of magnetized D9 branes in a
toroidal orbifold compactification of Type IIB string theory.

To set our notations, let us first give some details on the background geometry.
We take the space-time to be the product of
$\mathbb{R}^{1,3}$ times a six-dimensional factorized torus
$\mathcal{T}_6=\mathcal{T}_2^{(1)}
\times\mathcal{T}_2^{(2)}\times\mathcal{T}_2^{(3)}$. For each
torus $\mathcal{T}_2^{(i)}$, the string frame metric
and the $B$-field are parameterized by the K\"ahler and complex structure
moduli, respectively $T^{(i)}=T_1^{(i)}+\ii \,T_2^{(i)}$ and $U^{(i)}=U_1^{(i)}+
\ii \,U_2^{(i)}$, according to
\begin{equation}
\label{GB}
G^{(i)}
= \frac{T_2^{(i)}}{U_2^{(i)}}\,
\begin{pmatrix}1 & U_1^{(i)} \\ U_1^{(i)} &
|U^{(i)}|^2\end{pmatrix}~~~~\mbox{and}~~~~
B^{(i)}
= \begin{pmatrix}\,
0 & - T_1^{(i)} \\ T_1^{(i)} & 0
\end{pmatrix}~.
\end{equation}
In our conventions, the dimensionful volume of the $i$-th torus is
$(2\pi\sqrt{\alpha'})^2T_2^{(i)}$.
This toroidal geometry breaks $\mathrm{SO}(1,9)$ into
$\mathrm{SO}(1,3)\times \prod_i\mathrm{U}(1)^{(i)}$, and correspondingly the
ten-dimensional string coordinates $X^M$ and $\psi^M$ are split as
\begin{equation}
X^M \to (X^\mu, Z^i)
~~~~{\rm and}~~~~
\psi^M \to (\psi^\mu, \Psi^i)
\label{coordinates}
\end{equation}
where $\mu=0,1,2,3$ and~\footnote{The prefactors in (\ref{zipsii})
are chosen in such a way that the complex coordinates are
orthonormal in the metric (\ref{GB}).}
\begin{equation}
Z^i=\sqrt{\frac{T_2^{(i)}}{2U_2^{(i)}}}\left({X^{2i+2}+ U^{(i)}
X^{2i+3}}\right)~~,~~~
\Psi^i=\sqrt{\frac{T_2^{(i)}}{2U_2^{(i)}}}\left({\psi^{2i+2}+
U^{(i)}\psi^{2i+3}}\right)
\label{zipsii}
\end{equation}
for $i=1,2,3$.
Similarly, the (anti-chiral)%
\footnote{We define the 10-dimensional GSO projection so
that in the Ramond zero-mode sector it selects anti-chiral states; in other
words, in this sector we take $(-1)^F$ to be given by \emph{minus} the chirality
matrix $\Gamma_{11}$.}
spin-fields $S^{\dot{\mathcal{A}}}$ of the RNS formalism in ten dimensions
factorize in a product of four-dimensional and internal spin-fields according to
\begin{equation}
S^{\dot{\mathcal{A}}}
\to (S_\alpha S_{---},
S_\alpha S_{-++},S_\alpha S_{+-+},S_\alpha S_{++-},S^{\dot\alpha} S^{+++},
S^{\dot\alpha} S^{+--},S^{\dot\alpha} S^{-+-},S^{\dot\alpha} S^{--+})
\label{spin}
\end{equation}
where the index $\alpha$ ($\dot\alpha$) denotes positive (negative) chirality
in $\mathbb{R}^{1,3}$ and the labels $(\pm,\pm,\pm)$ on the
internal spin-fields denote charges
$(\pm\frac12,\pm\frac12,\pm\frac12)$ under the three internal
$\mathrm{U}(1)$'s.

Without loss of generality, we set the $B$-field to zero (at the end of this
section we will see how to incorporate it). The above geometry can also be
described in the so-called supergravity basis using the complex moduli $s$,
$t^{(i)}$ and $u^{(i)}$, whose relation with the  previously introduced
quantities in the string basis is (see for instance
Ref.~\cite{Lust:2004cx,Blumenhagen:2006ci})
\begin{equation}
\begin{aligned}
&\mathrm{Im}(s) \equiv s_2 = \frac{1}{4\pi}\,\ee^{-\phi_{10}}\,
T_2^{(1)}T_2^{(2)}T_2^{(3)}~,
\\
&\mathrm{Im}(t^{(i)}) \equiv t_2^{(i)} = \ee^{-\phi_{10}} T_2^{(i)}~,
\\
& u^{(i)} = u_1^{(i)} + \ii\, u_2^{(i)} = U^{(i)}~,
\end{aligned}
\label{stu}
\end{equation}
where $\phi_{10}$ is the ten dimensional dilaton. The real parts of $s$ and
$t^{(i)}$
are related to suitable R-R potentials.
In terms of these variables, the bulk K\"ahler potential in the $\mathcal{N}=1$ language%
\footnote{Strictly speaking this K\"ahler potential is not globally defined
since the scalars of the hypermultiplets $T^{(1)},T^{(2)} $ and
$U^{(1)},U^{(2)} $  live in a quaternionic manifold, which is not
K\"ahler since its holonomy group is not contained in $U(n)$.
The quaternionic manifold of $N=2$ supergravity becomes an
hyperK\"ahler manifold of $N=2$ rigid supersymmetry in the limit where
the gravitational interaction is switched off: the K\"ahler potential we use
in this work has therefore to be interpreted as the expression one
obtains in the rigid limit or as a local expression.
}
is given by
\cite{Antoniadis:1996vw}
\begin{equation}
K = -\log (s_2) -\sum_{i=1}^3 \log(t_2^{(i)}) -
\sum_{i=1}^3 \log(u_2^{(i)})~.
\label{kpot}
\end{equation}

\subsection{The gauge sector}
\label{subsec:gaugesector}

In the above toroidal background we now introduce a stack of $N_a$ D9 branes.
The open string excitations that are massless in
$\mathbb{R}^{1,3}$ describe a Super Yang-Mills
(SYM) theory with gauge group $\mathrm{U}(N_a)$ and $\mathcal{N}=4$
supersymmetry in four dimensions. In order to reduce to $\mathcal{N}=2$, we
replace
$\mathcal{T}_6$ with the toroidal orbifold
\begin{equation}
\frac{\mathcal{T}_2^{(1)}\times\mathcal{T}_2^{(2)}}{\mathbb{Z}_2}\times\mathcal{
T}_2^{(3)}~,
\label{orbifold}
\end{equation}
where $\mathbb{Z}_2$ simply acts as a reflection in the first two
tori ({\it i.e.} $Z^i\to-Z^i$ for $i=1,2$), and consider
fractional D9 branes instead of bulk branes%
\footnote{The twisted closed string sectors introduced by the orbifold
will not play any r\^ole for our considerations, and thus it is enough
to still consider only the  untwisted moduli (\ref{stu}).}. Actually, in the
orbifold (\ref{orbifold}) there are two types of fractional branes
corresponding to the two irreducible representations of
$\mathbb{Z}_2$ that can be assigned to the open string Chan-Paton factors.
For simplicity, we take all the $N_a$ D9 branes to be fractional
branes of the same kind (for example with the trivial
representation on the Chan-Paton factors) and we will call them
color branes. Then,
one can easily see that the physical massless open string states
surviving the orbifold projection are a vector $A_\mu$, a complex scalar $\phi$
and two gaugini $\Lambda^{\alpha1}$ and $\Lambda^{\alpha2}$.
They are described by the following vertex operators
\begin{subequations}
\begin{align}
&V_A(z) =(\pi\alpha')^{\frac12}\,{A_\mu}\,\psi^\mu(z)\,
\ee^{-\varphi(z)}\,\ee^{\ii
p_\mu X^\mu(z)}~,
\label{vertA}\\
&V_\phi(z) =(\pi\alpha')^{\frac12}\,{\phi}\,\Psi^3(z)\,
\ee^{-\varphi(z)}\,\ee^{\ii
p_\mu X^\mu(z)}
\label{vertphi}
\end{align}
\label{vertbos}
\end{subequations}
in the (--1) superghost picture of the NS sector, and
\begin{subequations}
\begin{align}
&V_{\Lambda^1}(z)
=(2\pi\alpha')^{\frac34}\,\Lambda^{\alpha1}\,S_{\alpha}(z)S_{+-+}(z)\,
\ee^{-\frac12\varphi(z)}\,\ee^{\ii
p_\mu X^\mu(z)}~,
\label{vertlam1}\\
&V_{\Lambda^2}(z)
=(2\pi\alpha')^{\frac34}\,\Lambda^{\alpha2}\,S_{\alpha}(z)S_{-++}(z)\,
\ee^{-\frac12\varphi(z)}\,\ee^{\ii
p_\mu X^\mu(z)}
\label{vertlam2}
\end{align}
\label{vertferm}
\end{subequations}
in the (--1/2) superghost picture of the R sector. We have defined
the action of the $\mathbb{Z}_2$ orbifold generator $h$ on the
R ground states to be
\begin{equation}
h = -\sigma_3 \otimes \sigma_3 \otimes 1~,
\label{gorb}
\end{equation}
which is the spinor representation of a $\pi$ rotation in the
first two tori. Then, one can easily see that the two internal spin fields
in the fermionic vertices (\ref{vertferm}) have $h$-parity one and
are selected by the orbifold projection
\begin{equation}
P_{\mathrm{orb}}= \frac{1+h}2~.
\label{projorb}
\end{equation}
In all vertices (\ref{vertbos}) and (\ref{vertferm}), the polarizations have
canonical
dimensions (this explains the dimensional prefactors
\footnote{See for example Ref. \cite{Billo:2002hm} for details
on the normalizations of vertex operators and scattering
amplitudes.}) and are $N_a\times N_a$ matrices transforming
in the adjoint representation of $\mathrm{SU}(N_a)$ (here we neglect an overall
factor of
$\mathrm{U}(1)$, associated to the center of mass of the $N_a$ D9 branes, which
decouples and
does not play any r\^ole in our present context).
The vertex operators (\ref{vertbos}) and (\ref{vertferm})
describe the components of a $\mathcal{N}=2$ vector superfield
and are connected to each other by the following
supercharges:
\begin{equation}
\begin{aligned}
Q_{\alpha 1}&= \oint \frac{dz}{2\pi{\rm i}} S_{\alpha}(z)S_{-++}(z)\,
{\rm e}^{-\frac12\varphi(z)}
~,~~
Q_{\alpha2}= \oint \frac{dz}{2\pi{\rm i}} S_{\alpha}(z)S_{+-+}(z)\,
{\rm e}^{-\frac12\varphi(z)}~,
\\
{\bar Q}^{\dot\alpha 1}&= \oint \frac{dz}{2\pi{\rm i}}
S^{\dot\alpha}(z)S^{+--}(z)\,
{\rm e}^{-\frac12\varphi(z)}
~,~~
{\bar Q}^{\dot\alpha 2}= \oint \frac{dz}{2\pi{\rm i}}
S^{\dot\alpha}(z)S^{-+-}(z)\,
{\rm e}^{-\frac12\varphi(z)}~,
\end{aligned}
\label{n2susy}
\end{equation}
which generate the $\mathcal{N}=2$ supersymmetry algebra selected by the
orbifold projection (\ref{projorb}).

By computing all tree-level
scattering amplitudes among the vertex operators
(\ref{vertbos}) and (\ref{vertferm}) and their conjugates,
and taking the field theory limit $\alpha'\to 0$,
one can obtain the ${\cal N}=2$ SYM action
\begin{equation}
\begin{aligned}
&S_{\rm SYM}=\frac{1}{g_a^2}\,
\int d^4x ~{\rm
Tr}\,\Big\{\frac{1}{2}\,F_{\mu\nu}^2 +2\, D_\mu \bar{\phi}\,D^\mu \phi
-2\,\bar\Lambda_{\dot\alpha A}\bar D\!\!\!\!/^{\,\dot\alpha \beta}
\Lambda_\beta^{\,A}\\
&~~~~~~~~+{\rm i}\sqrt 2 \,\bar\Lambda_{\dot\alpha A}\epsilon^{AB}
\big[\,\phi, \bar\Lambda^{\dot\alpha}_{\,B}\big]
+{\rm i}\sqrt 2\, \Lambda^{\alpha A}\epsilon_{AB}
\big[\,\bar \phi, \Lambda_{\alpha}^{\,B}\big]+\big[\,\phi,
\bar\phi\,\big]^2~\Big\}~,
\end{aligned}
\label{n2}
\end{equation}
where $A,B=1,2$, and the Yang-Mills coupling constant $g_a$ is given by
\begin{equation}
\frac{1}{g_a^2} = \frac{1}{4\pi}\,\ee^{-\phi_{10}}\,T_2^{(1)}T_2^{(2)}T_2^{(3)}
= s_2~.
\label{gym}
\end{equation}
Since we will study instanton effects, we have written the above action with
Euclidean signature.

For later convenience it is useful to compare the bosonic part of the action
(\ref{n2}) with the Dirac-Born-Infeld (DBI) action for D9 branes in the toroidal
orbifold (\ref{orbifold}). In the Euclidean string frame, this action is given
by
\begin{equation}
S_{\rm DBI} = \frac{2\pi}{(2\pi\sqrt{\alpha'})^{10}}\,
\int d^{10}x\,\ee^{-\phi_{10}}\,\sqrt{
\det\big(G_{MN}+2\pi\alpha'F_{MN}\big)}~,
\label{dbi}
\end{equation}
where $G_{MN}$ is the world-volume metric and $F_{MN}$ is a gauge field
strength.
Promoting the latter to be non-abelian~\footnote{We normalize the generators
$T_A$
of the gauge group such that
$\mathrm{Tr}\left(T_AT_B\right)=\frac{1}{2}\delta_{AB}$.},
and compactifying $S_{\rm DBI}$ to four dimensions
on the toroidal orbifold (\ref{orbifold}),
the quadratic terms in $F$ read (see also Refs. \cite{Berg:2005ja,Di
Vecchia:2005vm})
\begin{equation}
\int d^4x\,\sqrt{\det G_4}\,\,{\rm Tr}\Big\{
\frac{1}{2g_a^2}\,F_{\mu\nu}^2 +
2\,\ee^{-\phi_{10}}\prod_{i=1}^3
\sqrt{\det G^{(i)}}\frac{1}{T_2^{(3)}U_2^{(3)}}
\,D_\mu{\bar \Phi}D^\mu \Phi
\Big\}~,
\label{dbi1}
\end{equation}
where $G_4$ is the string frame metric in the non-compact space and
\begin{equation}
\Phi= \frac{1}{\sqrt{4\pi}}\left(U^{(3)} A_8-A_9\right)
\label{C}
\end{equation}
with $A_8$ and $A_9$ denoting the components of the ten-dimensional gauge field
along
$\mathcal{T}_2^{(3)}$. Changing to the (flat) Euclidean Einstein frame
with
\begin{equation}
(G_4)_{\mu\nu} = \ee^{2\phi_4}\,\delta_{\mu\nu}~,
\label{stringeinstein}
\end{equation}
where $\phi_4=\phi_{10} -\frac{1}{2}\sum_i \log(T_2^{(i)})$
is the four-dimensional dilaton, and using the geometrical moduli (\ref{stu})
of the supergravity basis, we can rewrite (\ref{dbi1}) as
\begin{equation}
\int d^4x\,{\rm Tr}\Big\{\frac{1}{2g_a^2}\,F_{\mu\nu}^2 + 2
\,K_\Phi\,D_\mu{\bar \Phi}D^\mu \Phi
\Big\}~,
\label{dbi2}
\end{equation}
where we have introduced the K\"ahler metric for
$\Phi$, namely
\begin{equation}
K_\Phi= \frac{1}{t_2^{(3)} u_2^{(3)}}~.
\label{zc}
\end{equation}
This K\"ahler metric can be obtained directly also from a 3-point scattering
amplitude involving one of the (closed string) geometric moduli and two scalar
fields, as explained for example in
Refs.~\cite{Bertolini:2005qh,Lust:2004cx}, after appropriate changes from the
string to the supergravity basis.

Comparing (\ref{dbi2}) with the bosonic kinetic terms in (\ref{n2}), we see that
the relation between the canonically normalized field $\phi$ appearing in the
string vertex operators and the field $\Phi$ in the supergravity basis is
\begin{equation}
\phi = g_a\sqrt{K_\Phi}\,\Phi~.
\label{phiC}
\end{equation}

\subsection{The matter sector}
\label{subsec:mattersector}

We now want to add $\mathcal{N}=2$ hyper-multiplets in this orbifold set up. The
simplest possibility to do this is to add a second stack of fractional D9 branes
(flavor branes) which carry a different representation of the orbifold group as
compared to the color branes considered so far. The massless open strings
stretching between the flavor branes and the color branes account precisely for
$\mathcal{N}=2$ hyper-multiplets in the fundamental representation of the gauge
group $\mathrm{SU}(N_a)$. However, we can be more general than this and
introduce {\it magnetized} flavor D9 branes. To distinguish them from the color
branes, we will denote their various parameters with a subscript $b$. For
example, $N_b$ will be their number and $n_b^{(i)}$ will be their wrapping
number around the $i$-th torus.

Introducing a magnetic flux on the $i$-th torus for the flavor branes amounts to
pick a $\mathrm{U}(1)$ subgroup in the Cartan subalgebra of
$\mathrm{U}(n_b^{(i)})$ and turn on a constant magnetic field%
\footnote{Even if more general magnetizations could be introduced, for simplicity we will consider only ``diagonal'' magnetic fields which respect the factorized structure of the internal toroidal space.} $F_b^{(i)}$,
namely
\begin{equation}
F_b^{(i)}= f_b^{(i)}\,dX^{2i+2}\wedge dX^{2i+3} =
\ii\,\frac{f_b^{(i)}}{T_2^{(i)}}\,
dZ^i\wedge d{\bar Z}^i =
\frac{f_b^{(i)}}{\sqrt{G^{(i)}}}\, J^{(i)}~,
\label{fi}
\end{equation}
where in the last step we have introduced the K\"ahler form $J^{(i)}$.
The generalized Dirac quantization condition requires that the
first Chern class $c_1(F_b^{(i)})$ be an integer, namely
\begin{equation}
c_1(F_b^{(i)})=\frac{1}{2\pi}\int_{\mathcal{T}_2^{(i)}}
\mathrm{Tr}(F_b^{(i)})= \frac{1}{2\pi} (2\pi\sqrt{\alpha'})^2 n_b^{(i)}
f_b^{(i)}
 =m_b^{(i)} \in
\mathbb{Z}~,
\label{c1}
\end{equation}
that is
\begin{equation}
2\pi\alpha' f_b^{(i)} = \frac{m_b^{(i)}}{n_b^{(i)}}~.
\label{nm}
\end{equation}
The total magnetic field is then
$F_b=F_b^{(1)}+F_b^{(2)}+
F_b^{(3)}$. In order to preserve at least $\mathcal{N}=1$
supersymmetry in the bulk, the magnetic field has to satisfy the
relation
\begin{equation}
J\wedge J\wedge \hat F_b= \frac{1}{3} \,\hat F_b\wedge \hat F_b\wedge \hat
F_b~,
\label{fwedgef}
\end{equation}
where $\hat F_b=2\pi\alpha' F_b$ and $J$ is the total K\"ahler form
$J=\sum_i J^{(i)}$.
Setting
\begin{equation}
2\pi\alpha' \frac{f_b^{(i)}}{T_2^{(i)}} = \tan \pi\nu_b^{(i)}~~~~{\rm
with}~~~~0\leq \nu_b^{(i)}
< 1~,
\label{nui}
\end{equation}
it is easy to see that the supersymmetry requirement (\ref{fwedgef}) is
fulfilled if \footnote{Other solutions of (\ref{fwedgef}) are
$-\nu_b^{(1)}-\nu_b^{(2)}+\nu_b^{(3)}=0$;
$-\nu_b^{(1)}+\nu_b^{(2)}-\nu_b^{(3)}=0$;
$\nu_b^{(1)}+\nu_b^{(2)}+\nu_b^{(3)}=2$. They are all related to the
solution (\ref{nu123}) by obvious changes.}
\begin{equation}
\nu_b^{(1)}-\nu_b^{(2)}-\nu_b^{(3)}=0~.
\label{nu123}
\end{equation}
If we want to have the same $\mathcal{N}=2$ supersymmetry which is
realized by the orbifold (\ref{orbifold}), we have to set
\begin{equation}
\nu_b^{(3)}=0\quad\mbox{and hence}\quad\nu_b^{(1)}=\nu_b^{(2)}~.
\label{n12}
\end{equation}
This implies that the open strings stretching between the flavor
branes and the color branes ({\it i.e.} the D$9_b$/D$9_a$ strings)
are twisted only along the directions of the first two tori. More
specifically, the internal string coordinates $Z^i$ and $\Psi^i$ defined in
(\ref{zipsii}) satisfy, for $i=1,2$, the following twisted monodromy properties
\begin{equation}
Z^i\big({\rm e}^{2\pi{\rm i}} z\big)= \,{\rm e}^{2\pi{\rm
i}\nu^{(i)}_{b}}\,Z^i(z)~~~{\mbox{and}}~~~
\Psi^i\big({\rm e}^{2\pi{\rm i}} z\big)= \eta\,{\rm e}^{2\pi{\rm
i}\nu^{(i)}_{b}}\,\Psi^i(z)~,
\label{monodromy}
\end{equation}
where $\eta=+1$ for the NS sector and $\eta=-1$ for the R sector. On the
other hand, $Z^3$ and $\Psi^3$ have the usual untwisted properties.

Let us now describe the physical massless states of the D$9_b$/D$9_a$
strings, starting from the NS sector.
To write the vertex operators it is convenient to introduce the following
notation
\begin{equation}
\sigma(z)\equiv\prod_{i=1}^2\sigma_{\nu_{b}^{(i)}}(z)~,\quad
s(z)\equiv\prod_{i=1}^2S_{\nu_{b}^{(i)}}~,
\label{sigma}
\end{equation}
where $\sigma_{\nu_{b}^{(i)}}$ and $S_{\nu_{b}^{(i)}}$ are respectively
the bosonic and fermionic twist fields in the
$i$-th torus whose conformal dimensions are
\begin{equation}
h_{\sigma}^{(i)}= \frac{1}{2}\nu_{b}^{(i)}\big(1-\nu_{b}^{(i)}\big)
~~~{\mbox{and}}~~~h_{S}^{(i)}= \frac{1}{2}\big(\nu_{b}^{(i)}\big)^2~.
\label{h}
\end{equation}
Then, the physical massless states are described by the
following vertex operators:
\begin{equation}
\begin{aligned}
V_{q}(z) &= (2\pi\alpha')^{\frac12}\,q\,\,
\sigma(z):\!\bar\Psi^1(z)s(z)\!:
{\rm e}^{-\varphi(z)}\,{\rm e}^{{\rm i}p_\mu X^\mu(z)}~,
\\
V_{{\tilde {q}}^\dagger}(z) &= (2\pi\alpha')^{\frac12}\,{\tilde
{q}}^\dagger\,\,
\sigma(z):\!\bar\Psi^2(z)s(z)\!:
{\rm e}^{-\varphi(z)}\,{\rm e}^{{\rm i}p_\mu X^\mu(z)}~,
\label{vertn2scal}
\end{aligned}
\end{equation}
which can be easily checked to have conformal dimension 1 for
$p^2=0$ if $\nu^{(1)}_{b}=\nu^{(2)}_{b}$.

In the R sector instead the massless states are described by the following
vertex operators
\begin{equation}
\begin{aligned}
V_{\chi}(z) &= (2\pi\alpha')^{\frac34}\,
{\chi}^{\alpha}\,S_{\alpha}(z)\,\sigma(z)\Sigma(z)S_{-}(z)
\,
{\rm e}^{-\frac12\varphi(z)}\,{\rm e}^{{\rm i}p_\mu X^\mu(z)}~,
\\
V_{{\tilde \chi}^\dagger}(z) &=
(2\pi\alpha')^{\frac34}\,{\tilde\chi}^\dagger_{\dot\alpha}\,
\,S^{\dot\alpha}(z)\,\sigma(z)\Sigma(z)
S_{+}(z)\,
{\rm e}^{-\frac12\varphi(z)}\,{\rm e}^{{\rm i}p_\mu X^\mu(z)}~,
\end{aligned}
\label{vertn2ferm}
\end{equation}
where
\begin{equation}
\Sigma(z)=\prod_{i=1}^2S_{\nu_{b}^{(i)}-\frac12}(z)
\label{Sigma}
\end{equation}
and $S_{\pm}$ are the spin fields in the untwisted directions of the third
torus. Again, one can easily check that these vertex operators have conformal
dimension 1 for $p^2=0$.

In all the above vertices the polarizations, which carry a color index in the
fundamental representation of $\mathrm{SU}(N_a)$, have canonical dimensions and
are odd under $\mathbb{Z}_2$, since they describe open strings that connect
fractional branes belonging to different irreducible representations of the
orbifold group. Consequently we must require that the operator part in
(\ref{vertn2scal}) and (\ref{vertn2ferm}) be also odd under $\mathbb{Z}_2$ so
that altogether the complete vertices can survive the orbifold projection. In
particular this implies that the twisted part of the R ground states, described
by $\sigma(z)\Sigma(z)$, must be declared odd under $\mathbb{Z}_2$ while the
twisted part of the NS ground states, described by $\sigma(z)s(z)$, must be
declared even. The vertices (\ref{vertn2scal}) and (\ref{vertn2ferm}) are
connected to each other by the same eight supercharges (\ref{n2susy}) that are
selected by the $\mathbb{Z}_2$ orbifold, and thus their polarizations form a
hyper-multiplet representation of $\mathcal{N}=2$ supersymmetry. More precisely,
taking into account the multiplicity of the $(a,b)$ intersection, they can be
organized into $N_F$ hyper-multiplets whose components in the following will be
denoted as $\big(q_f,{\tilde q_f}^\dagger,{\chi}_f,{\tilde \chi_f}^\dagger\big)$
with $f=1,\ldots,N_F$.

The D$9_a$/D$9_b$ strings with opposite orientation have a completely similar
structure; at the massless level the physical vertex operators are
\begin{equation}
\begin{aligned}
V_{q^\dagger}(z) &= (2\pi\alpha')^{\frac12}\,q^\dagger\,\,
\bar\sigma(z):\!\Psi^1(z)\bar s(z)\!:
{\rm e}^{-\varphi(z)}\,{\rm e}^{{\rm i}p_\mu X^\mu(z)}~,
\\
V_{\tilde q}(z) &= (2\pi\alpha')^{\frac12}\,{\tilde q}\,\,
\bar\sigma(z):\!\Psi^2(z)\bar s(z)\!:
{\rm e}^{-\varphi(z)}\,{\rm e}^{{\rm i}p_\mu X^\mu(z)}
\label{vertn2scal2}
\end{aligned}
\end{equation}
in the NS sector, and
\begin{equation}
\begin{aligned}
V_{\chi^\dagger}(z) &= (2\pi\alpha')^{\frac34}\,
{\chi^\dagger_{\dot\alpha}}\,\,S^{\dot\alpha}(z)\,\bar\sigma(z)
\bar\Sigma(z)S_{+}(z)
\,
{\rm e}^{-\frac12\varphi(z)}\,{\rm e}^{{\rm i}p_\mu X^\mu(z)}~,
\\
V_{\tilde \chi}(z) &=
(2\pi\alpha')^{\frac34}\,{\tilde \chi}^{\alpha}\,\,S_{\alpha}(z)
\,\bar\sigma(z)\bar\Sigma(z)S_{-}(z)\,
{\rm e}^{-\frac12\varphi(z)}\,{\rm e}^{{\rm i}p_\mu X^\mu(z)}
\end{aligned}
\label{vertn2ferm2}
\end{equation}
in the R sector. Here we have defined the anti-twist fields as follows:
\begin{equation}
\label{twistdef}
\bar\sigma(z)\equiv\prod_{i=1}^2\sigma_{1-\nu_{b}^{(i)}}(z)~,\quad
\bar s(z)\equiv\prod_{i=1}^2S_{-\nu_{b}^{(i)}}(z)
~,\quad
\bar\Sigma(z)\equiv\prod_{i=1}^2S_{\frac12-\nu_{b}^{(i)}}(z)~.
\end{equation}
The vertices (\ref{vertn2scal2}) and (\ref{vertn2ferm2}) are conjugate
to the ones in (\ref{vertn2scal}) and (\ref{vertn2ferm}) respectively.

By computing all tree-level scattering amplitudes among the above vertex
operators and those of gauge sector, and taking the field theory limit
$\alpha'\to 0$, one can obtain the $\mathcal{N}=2$ action for hyper-multiplets
coupled to a vector multiplet. For example, from the computation of a 3-point
function between a gluon, a scalar of the hyper-multiplet and its conjugate, one
can reconstruct the kinetic terms
\begin{equation}
\int d^4x ~\sum_{f=1}^{N_F}\Big\{D_\mu{q^\dagger}^f\,D^\mu {q}_f
+ D_\mu {\tilde q}^f\,D^\mu {{\tilde q}^\dagger}_f \Big\}~,
\label{kinq}
\end{equation}
where we have explicitly indicated the sum over the flavor indices and
suppressed the color indices. Similarly, from other 3-point functions one can
obtain the various Yukawa interactions, like for example
\begin{equation}
\int d^4x ~\sum_{f=1}^{N_F} \,{\tilde\chi}^f\,\phi\,\chi_f~.
\label{yuk}
\end{equation}

In the supergravity basis it is customary to use fields with a different
normalization and write for example the kinetic term for the scalars of the
hyper-multiplet as
\begin{equation}
\int d^4x ~\sum_{f=1}^{N_F}K_Q\Big\{ D_\mu {Q^\dagger}^f\,D^\mu {Q}_f
+ D_\mu {\tilde Q}^f\,D^\mu {\tilde Q}^\dagger_f \Big\}~.
\label{kinQ}
\end{equation}
Upon comparison with
(\ref{kinq}), we see that the relation between the canonically normalized fields
$q$ and $\tilde q$ appearing in the string vertex operators and the fields $Q$
and $\tilde Q$ of the supergravity basis is
\begin{equation}
q =\sqrt{K_Q}\,Q\quad\mbox{and}\quad
\tilde q = \sqrt{K_Q}\,\tilde Q ~.
\label{qQ}
\end{equation}
On the other hand, using a $\mathcal{N}=1$ language in the supergravity basis,
the various Yukawa couplings can be encoded in the holomorphic superpotential
\begin{equation}
W= \sum_{f=1}^{N_F} \,{\tilde Q}^f \,\Phi\,Q_f~,
\label{wyuk}
\end{equation}
where we have adopted for the chiral superfields the same notation used for
their bosonic components.

By explicitly writing  the relation between the Yukawa couplings in the
canonical basis (see {\it e.g.} (\ref{yuk})) and those in the supergravity basis
derived from the $\mathcal{N}=1$ superpotential (\ref{wyuk}), we obtain
\begin{equation}
1 = \ee^{{K}/2}\, \big(\sqrt{K_Q}\big)^{-2}\,
\big(g_a\,\sqrt{K_\Phi}\big)^{-1}~,
\label{rel1}
\end{equation}
where the factor $\ee^{{K}/2}$ is the contribution of the bulk supergravity
K\"ahler potential. Clearly, we can rewrite (\ref{rel1}) also as
\begin{equation}
\ee^{{K}/2}\,K_Q^{-1}= g_a\sqrt{K_\Phi}~,
\label{rel2}
\end{equation}
which will be useful later \footnote{In Section \ref{sec:concl} we will rewrite this
relation in a full fledged $\mathcal{N}=2$ notation (see Eq. (\ref{rel22})).}.
Using (\ref{gym}), the expression for the K\"ahler
potential $K$ given in (\ref{kpot}) and the K\"ahler metric $K_\Phi$ given in
(\ref{zc}), we deduce that
\begin{equation}
K_Q= \frac{1}{\big(t_2^{(1)}t_2^{(2)}u_2^{(1)}u_2^{(2)}\big)^{1/2}}~.
\label{kQ}
\end{equation}
This expression for $K_Q$ agrees with the one mentioned in Ref. \cite{Akerblom:2007uc}.
It is worth pointing out that also the metric (\ref{kQ}) can be reconstructed from a 3-point
scattering amplitude along the lines discussed in Refs. \cite{Bertolini:2005qh,Lust:2004cx},
after the appropriate changes between the string and the supergravity basis are
taken into account.

In the following we will consider also the case in which the hyper-multiplets
are massive with a $\mathcal{N}=2$ invariant mass term given by
\begin{equation}
W_M = \sum_{f=1}^{N_F}\, M_f\,{\tilde Q}^f\,Q_f~.
\label{mass}
\end{equation}
{F}rom this superpotential we immediately see that the corresponding mass
parameters $m_f$ appearing in the canonically normalized action of the string
basis are
\begin{equation}
m_f = \ee^{{K}/2}\,K_Q^{-1}\, M_f = g_a\sqrt{K_\Phi}\,M_f~,
\label{mf}
\end{equation}
where in the last step we have used (\ref{rel2}). Comparing this expression with
(\ref{phiC}), we see that $m_f$ and $\phi$ are related to the corresponding
quantities $M_f$ and $\Phi$ in the supergravity basis in the same way.

\subsection{Generalizations}
\label{subsec:generalizations}
The above construction can be easily generalized in several ways. For example,
if a background $B$ field is turned on in the internal space, see (\ref{GB}),
the magnetic flux $2\pi\alpha' f_b^{(i)}$ gets replaced by
$\widehat{f}_b^{(i)}\equiv 2\pi\alpha' f_b^{(i)}-T_1^{(i)}$, so that (\ref{nui})
becomes
\begin{equation}
\tan \pi\nu_b^{(i)}= \frac{m_b^{(i)}-n_b^{(i)}T_1^{(i)}}{n_b^{(i)}T_2^{(i)}}~,
\label{nuib}
\end{equation}
where the quantization condition (\ref{nm}) has been taken into account.
Note that in the presence of $B$ also the color branes acquire intrinsic
twist parameters given by
\begin{equation}
\tan \pi\nu_a^{(i)}= -\frac{T_1^{(i)}}{T_2^{(i)}}
\label{nuia0}
\end{equation}
and the monodromy properties of the D$9_b$/D$9_a$ strings
depend on the relative twist parameters
\begin{equation}
\nu_{ba}^{(i)}=\nu_{b}^{(i)}-\nu_{a}^{(i)}
\label{nuab}
\end{equation}
which must replace $\nu_{b}^{(i)}$ in the various vertex operators like
(\ref{vertn2scal}) and (\ref{vertn2ferm}). We can further generalize this by
wrapping the color branes $n_a^{(i)}$ times on the $i$-th torus and turning on a
magnetic field on their world volume with integer magnetic numbers $m_a^{(i)}$.
In this way the intrinsic twist parameters $\nu_a^{(i)}$ of the color branes
have the same expression as (\ref{nuib}) with the subscript $b$ replaced by $a$.

Non-trivial wrapping and magnetic numbers for the color branes also influence
the explicit expressions of the various quantities in the effective gauge
theory. For example, the gauge coupling constant $g_a$ turns out to be given by
\begin{equation}
\frac{1}{g_a^2} =
\frac{1}{4\pi}\,\ee^{-\phi_{10}}\,\prod_{i=1}^3 \big|
n_a^{(i)} T^{(i)} - m_a^{(i)} \big|
=s_2
\,\left|\ell_a^{(1)}\ell_a^{(2)}\ell_a^{(3)}\right|~,
\label{gyma}
\end{equation}
where we have defined
\begin{equation}
\ell_a^{(i)} = \frac{n_a^{(i)}T^{(i)}-m_a^{(i)}}{T_2^{(i)}}~.
\label{vai}
\end{equation}
Note that if we use the supersymmetry relation
\begin{equation}
\prod_i \frac{\widehat{f}_a^{(i)}}{T_2^{(i)}}
=\sum_i \frac{\widehat{f}_a^{(i)}}{T_2^{(i)}}
\label{susy3}
\end{equation}
for the quantities $\widehat{f}_a^{(i)}\equiv 2\pi\alpha' f_a^{(i)}-T_1^{(i)}$,
which follows from the obvious extension of (\ref{fwedgef}), we can rewrite
(\ref{gyma}) as
\begin{equation}
\frac{1}{g_a^2}= n_a^{(1)}n_a^{(2)}n_a^{(3)}\,\Big|s_2 - \frac{1}{4\pi}\big(
\widehat{f}_a^{(1)}\widehat{f}_a^{(2)}t_2^{(3)}
+\widehat{f}_a^{(2)}\widehat{f}_a^{(3)}t_2^{(1)}+
\widehat{f}_a^{(3)}\widehat{f}_a^{(1)}t_2^{(2)}\big)
\Big|~.
\end{equation}

By repeating the same analysis of the previous subsections when the color branes
are magnetized, one finds that the K\"ahler metric for the adjoint scalar field
$\Phi$ is
\begin{equation}
K_\Phi = \frac{1}{t_2^{(3)}
u_2^{(3)}}\,\left|\frac{\ell_a^{(1)}\ell_a^{(2)}}{\ell_a^{(3)}}\right|
\label{zc1}
\end{equation}
and that the K\"ahler metric for the fundamental chiral multiplets
$Q$ and $\tilde Q$ is
\begin{equation}
K_Q=
\frac{1}{\big(t_2^{(1)}t_2^{(2)}u_2^{(1)}u_2^{(2)}\big)^{1/2}}\,\left|\ell_a^{
(3)}\right|~.
\label{kQ1}
\end{equation}
These expressions reduce to those given respectively in (\ref{zc}) and
(\ref{kQ}) when the color branes are not magnetized and are trivially wrapped on
the internal space, since in this case $|\ell_a^{(i)}|\to 1$ for all $i$.

Performing a T-duality transformation
\begin{equation}
T^{(i)}\to -\frac{1}{U^{(i)}}\quad,\quad U^{(i)}\to -\frac{1}{T^{(i)}}
\label{tduality}
\end{equation}
with the four-dimensional dilaton $\phi_4$ kept fixed, we can translate our
results for magnetized D$9$ branes into those for intersecting D6 branes of the
type IIA theory. Under this transformation, $t_2^{(i)}$ and $u_2^{(i)}$ are
interchanged, while
\begin{equation}
\ell_a^{(i)} \to -\frac{{\bar
U}^{(i)}}{U_2^{(i)}}\,\big(n_a^{(i)}+U^{(i)}m_a^{(i)}\big)~.
\label{ell1}
\end{equation}
We can therefore see that, after T-duality, the K\"ahler metrics (\ref{zc1}) and
(\ref{kQ1}) are a generalization of those presented in Ref.
\cite{Akerblom:2007uc} for intersecting D branes on rectangular tori ({\it i.e.}
$U_1^{(i)}=0$). Notice also that when all branes are magnetized, the number
$N_F$ of fundamental hyper-multiplets associated to the strings stretching
between the D$9_b$ and the D$9_a$ branes is given by
\begin{equation}
N_F=N_b\,I_{ba}=N_b\,I_{ab}~,
\label{nfl}
\end{equation}
where
\begin{equation}
I_{ab} = \prod_{i=1}^2\big(m_a^{(i)}n_b^{(i)}-
m_b^{(i)}n_a^{(i)}\big)=I_{ba}
\label{iab}
\end{equation}
represents the number of Landau levels for the $(a,b)$ intersection.

We finally observe that in a generic toroidal orbifold compactification with
wrapped branes there are unphysical closed string tadpoles that must be canceled
to have a globally consistent model. Usually this cancellation is achieved by
introducing an orientifold projection and suitable orientifold planes. Like in
other cases treated in the literature, in this paper we take a ``local'' point
of view focusing only on some intersections and assume that the model can be
made fully consistent with the orientifold projection.

\section{$\mathcal{N}=2$ instanton calculus from the string perspective}
\label{sec:inst_calc}
We now consider instanton effects in the $\mathcal{N}=2$ gauge theories
presented in the previous section. In this stringy set-up instanton
contributions can be obtained by adding fractional Euclidean D5 branes (nowadays
called E5 branes) that completely wrap the internal manifold $\frac{{\cal
T}_2^{(1)} \times {\cal T}_2^{(2)}}{{\mathbb Z}_2} \times {\cal T}_2^{(3)}$, and
hence describe point-like configurations from the four-dimensional point of
view. In general these E5 branes can be chosen with a representation of the
$\mathbb{Z}_2$ orbifold group on the Chan-Paton factors and/or with magnetic
fluxes that are different from the ones of the color D$9_a$ branes. If that is
the case, then the E5 branes represent ``exotic'' instantons whose properties
are different from those of the ordinary gauge theory instantons. Recently,
these ``exotic'' configurations  have been the subject of active investigations
\cite{Blumenhagen:2006xt}--\cite{Aharony:2007pr} from several different points
of view. Here we start by considering E5 branes that have the same
characteristics of the color D$9_a$ branes, except for their dimensions.
Therefore, we call them E$5_a$ branes. As we will see in detail later, these
E$5_a$ branes represent ordinary gauge instantons for the SYM theory on the
D$9_a$ branes. However, they are ``exotic'' instantons with respect to the gauge
theory defined on the flavor D$9_b$, and thus our results can be useful also for
the new developments.

The addition of $k$ E$5_a$ branes introduces new types of
excitations associated to open strings with at least one end-point on the
instantonic branes, namely the E$5_a$/E$5_a$ strings, the D$9_a$/E$5_a$ (or
E$5_a$/D$9_a$) strings and the D$9_b$/E$5_a$ (or E$5_a$/D$9_b$) strings. In all
these instantonic sectors, due to the Dirichlet-Dirichlet or mixed
Dirichlet-Neumann boundary conditions in the four non-compact directions, the
open string excitations do not carry any momentum and hence represent moduli
rather than dynamical fields in space-time. They however can carry (discretized)
momentum along the compact directions. Therefore we can distinguish the open
string states into those which do not carry any momentum in any directions and
those which do. The lightest excitations of the first type are truly instanton
moduli while those of the second type represent genuine string corrections whose
relevance for the effective theory will be elucidated in the following.

\subsection{Instanton moduli}
\label{subsec:1}
We now briefly list the instanton moduli for our $\mathcal{N}=2$
model which we distinguish into neutral, charged and flavored ones.

\paragraph{The neutral instanton sector}  The neutral instanton sector comprises
the zero-modes of open strings with both ends on the E$5_a$ branes. These modes
are usually referred to as neutral because they do not transform under the gauge
group. In the NS sector, after the $\mathbb{Z}_2$ orbifold projection, we find
six physical bosonic excitations that can be conveniently organized in a vector
$a_\mu$ and a complex scalar $\chi$, and also three auxiliary excitations $D_c$
($c=1,2,3$). The corresponding vertex operators are
\begin{subequations}
\label{vertNS}
\begin{align}
\label{verta}
&V_{a}(z)= g_{5_a}\,(2\pi\alpha')^{\frac12}\,a_\mu
\,\psi^{\mu}(z)\,\ee^{-\varphi(z)}~,
\\
\label{vertchi} &V_{\chi}(z)= \chi\,(\pi\alpha')^{\frac12}\,
{\Psi}^3(z)\,\ee^{-\varphi(z)}~,
\\
\label{vertd} &V_D(z)= {D_c}\,(\pi\alpha')\,\bar\eta_{\mu\nu}^c\,
\psi^\nu(z)\psi^\mu(z)~,
\end{align}
\end{subequations}
where $\bar\eta^c_{\mu\nu}$ are the three anti-self-dual 't Hooft
symbols and $g_{5_a}$ is the (dimensionful) coupling constant
on the E$5_a$, namely
\begin{equation}
\label{g0gym}
g_{5_a} = \frac{g_a}{4\pi^2\alpha'}
\end{equation}
with $g_a$ given in (\ref{gyma}). In the R sector, after the orbifold projection
(\ref{projorb}), we find four chiral fermionic zero-modes $M^{\alpha A}$
described by the vertex operators
\begin{equation}
\label{vertR1}
\begin{aligned}
V_{M^1}(z) &= \frac{g_{5_a}}{\sqrt 2}\,(2\pi\alpha')^{\frac34}\,
{M}^{\alpha 1}\,S_{\alpha}(z) S_{+-+}(z)\,\ee^{-\frac{1}{2}\varphi(z)}~,\\
V_{M^2}(z) &= \frac{g_{5_a}}{\sqrt 2}\,(2\pi\alpha')^{\frac34}\,
{M}^{\alpha 2} \,S_{\alpha}(z)
S_{-++}(z)\,\ee^{-\frac{1}{2}\varphi(z)}~,
\end{aligned}
\end{equation}
and four anti-chiral zero-modes $\lambda_{\dot\alpha A}$,
described by the vertices
\begin{equation}
\label{vertR2}
\begin{aligned}
V_{\lambda_1}(z)& = {{\lambda_{\dot\alpha
1}}}\,(2\pi\alpha')^{\frac34}\,S^{\dot\alpha}(z)S^{+--}(z)
\,\ee^{-\frac{1}{2}\varphi(z)}~,\\
V_{\lambda_2}(z)& = {{\lambda_{\dot\alpha
2}}}\,(2\pi\alpha')^{\frac34}\,S^{\dot\alpha}(z)S^{-+-}(z)
\,\ee^{-\frac{1}{2}\varphi(z)}~.
\end{aligned}
\end{equation}
All polarizations in the vertex operators (\ref{vertNS}), (\ref{vertR1}) and
(\ref{vertR2}) are $k\times k$ matrices and transform in the adjoint
representation of $\mathrm{U}(k)$. It is worth noticing that if the Yang-Mills
coupling constant $g_a$ is kept fixed when $\alpha'\to 0$, then the dimensionful
coupling $g_{5_a}$ in (\ref{g0gym}) blows up. Thus, some of the vertex operators
have been rescaled with factors of $g_{5_a}$ (like in (\ref{verta}) and
(\ref{vertR1})) in order to yield non-trivial interactions when $\alpha'\to 0$
\cite{Billo:2002hm}. As a consequence of this rescaling some of the moduli
acquire unconventional scaling dimensions which, however, are the right ones for
their interpretation as parameters of an instanton solution
\cite{Dorey:2002ik,Billo:2002hm}. For instance, the $a_\mu$'s have dimensions of
(length) and are related to the positions of the (multi-)centers of the
instanton, while ${M}^{\alpha A}$ have dimensions of (length)$^{\frac{1}{2}}$
and are the fermionic partners of the instanton centers. Furthermore, if we
write the $k\times k$ matrices ${a}^\mu$ and ${M}^{\alpha A}$ as
\begin{equation}
{a}^\mu = x_0^\mu\,\uno_{k\times k} + y^\mu_c\,T^c\quad,\quad
{M}^{\alpha A}=\theta^{\alpha A}\,\uno_{k \times k } + {\zeta}^{\alpha
A}_c\,T^c~,
\label{xtheta}
\end{equation}
where $T^c$ are the generators of $\mathrm{SU}(k)$, then the instanton center of
mass, $x_0^\mu$, and its fermionic partners, $\theta^{\alpha A}$, can be
identified respectively with the bosonic and fermionic coordinates of the
$\mathcal{N}=2$ superspace.

\paragraph{The charged instanton sector}  The charged instanton sector contains
the zero-modes of the open strings stretching between the color D$9_a$ branes
and the E$5_a$ branes, which transform in the fundamental representation of the
gauge group. In the NS sector there are two physical bosonic moduli
$w_{\dot\alpha}$ with dimension of (length) whose vertex operator is
\begin{equation}
\label{vertw}
V_w(z)= \frac{g_{5_a}}{\sqrt{2}}\,(2\pi\alpha')^{\frac12}\,{w}_{\dot\alpha}\,
\Delta(z) S^{\dot\alpha}(z)\,\ee^{-\varphi(z)}~.
\end{equation}
Here $\Delta$ is the twist operator with conformal weight $1/4$ which changes
the boundary conditions of the uncompact coordinates $X^\mu$ from Neumann to
Dirichlet. In the R sector there are two fermionic moduli $\mu^A$ with dimension
of (length)$^{1/2}$ whose vertices are
\begin{equation}
\begin{aligned}
V_{\mu^1}(z) &= \frac{g_{5_a}}{\sqrt{2}}\,
(2\pi\alpha')^{\frac34}\,{\mu}^1\, \Delta(z) S_{+-+}(z)\,
\ee^{-{\frac12}\varphi(z)}~,
\\
V_{\mu^2}(z)&= \frac{g_{5_a}}{\sqrt{2}}\,(2\pi\alpha')^{\frac34}\,
{\mu}^2\, \Delta(z) S_{-++}(z)\,
\ee^{-{\frac12}\varphi(z)}~.
\end{aligned}
\label{vertmu}
\end{equation}
Both in (\ref{vertw}) and (\ref{vertmu}) the polarizations are $N_a\times k$
matrices which transform in the bi-fundamental representation $(N_a,\bar k)$ of
${{\mathrm U}(N_a)\times\mathrm U}(k)$. Notice that these vertex operators are
even under the $\mathbb{Z}_2$ orbifold projection (\ref{projorb}). The charged
moduli associated to the open strings stretching from the E$5_a$ branes to the
D$9_a$'s, denoted by $\bar w_{\dot\alpha}$ and ${\bar\mu}^A$, transform in the
$(\bar N_a, k)$ representation  and are described by vertex operators of the same
form as (\ref{vertw}) and (\ref{vertmu}) except for the replacement of
$\Delta(z)$ by the anti-twist $\bar\Delta(z)$, corresponding to mixed
Dirichlet-Neumann boundary conditions along the four space-time directions. It
is worth pointing out that ${\bar\mu}^A$ are {\it not} the conjugates of
${\mu}^A$. This fact has important consequences for our purposes, as we will
discuss in Sect. \ref{sec:mix_ann}.

\paragraph{The flavored instanton sector} The flavored instanton sector
corresponds to the open strings that stretch between the flavor D$9_b$ branes
and the E$5_a$ branes. In this case the four non-compact directions have mixed
Neumann-Dirichlet boundary conditions while the complex coordinates along the
first two tori are twisted with parameters $\nu_{ba}^{(1)}=\nu_{ba}^{(2)}$ due
to the different magnetic fluxes at two end-points. As a consequence of this,
there are no bosonic physical zero-modes in the NS sector and the only physical
excitations are fermionic moduli with dimension of (length)$^{\frac12}$ from the
R sector, whose vertices are given by
\begin{equation}
\label{vertmup}
V_{\mu'}(z)= \frac{g_{5_a}}{\sqrt{2}}\,(2\pi\alpha')^{\frac34}\,
\mu'\, \Delta(z) \,\sigma(z)\Sigma(z)\,S_{-}(z)
\, \ee^{-{\frac12}\varphi(z)}~.
\end{equation}
Notice that this vertex operator is even under the $\mathbb{Z}_2$ orbifold
group, since both the operator part and the polarization are odd under
$\mathbb{Z}_2$, in complete analogy to what happens to the fermionic vertices
(\ref{vertn2ferm2}) of the flavored matter. Finally, we recall that the
zero-modes for the E$5_a$/D$9_b$ open strings with opposite orientation are
described by the vertex operators
\begin{equation}
\label{vertmup1}
V_{{\bar\mu}'}(z)= \frac{g_{5_a}}{\sqrt{2}}\,(2\pi\alpha')^{\frac34}\,
{\bar\mu}'\, \bar\Delta(z) \,\bar\sigma(z)\bar\Sigma(z)\,S_{-}(z)
\, \ee^{-{\frac12}\varphi(z)}~.
\end{equation}
Taking into account the multiplicity of the $(a,b)$ intersection, we will have
altogether $N_F$ fermionic moduli of each type which will be denoted
as $\mu'_f$ and ${{\bar\mu}'}{}^f$ with $f=1,\ldots,N_F$.

\smallskip
The physical moduli we have listed above, collectively called $\mathcal{M}_k$,
are in one-to-one correspondence with the ADHM moduli of $\mathcal{N}=2$ gauge
instantons (for a more detailed discussion see, for instance, Ref.
\cite{Dorey:2002ik} and references therein). In all instantonic sectors we can
construct many other open string states that carry a discretized momentum along
the compact directions and/or have some bosonic or fermionic string oscillators.
All these ``massive" states, however, are not physical, {\it i.e.} they cannot
be described by vertex operators of conformal dimension one, but, as we will see
later, they can play a role as internal states circulating in open string loop
diagrams.

\subsection{Instanton partition function}
Having identified the ADHM moduli, in analogy with the instanton calculus in
field theory we define the $k$-instanton partition function as the
``functional'' integral over the instanton moduli, namely
\begin{equation}
Z_{k}\,= \, {\cal C}_k\int d{\cal M}_k~\ee^{ -S ({\cal M}_k)}~,
\label{Z1}
\end{equation}
where ${\cal C}_k$ is a dimensional normalization factor which compensates for
the dimensions of the integration measure $d{\cal M}_k$, and $S({\cal M}_k)$ is
the moduli effective action which accounts for all possible interactions among
the instanton moduli in the limit $\alpha'\to 0$ (with $g_a$ fixed) at any order
of string perturbation theory. This action can be obtained by computing the
field theory limit of all scattering amplitudes with the vertex operators of the
ADHM moduli inserted on boundaries of open string world-sheets of any topology.
Formally  we can write
\begin{equation}
\begin{aligned}
- S({\cal M}_k) &= \sum_{\mathrm{topology}}
\langle\,1\,\rangle_{\mathrm{topology}}+
 \langle\,{\cal M}_k\,\rangle_{\mathrm{topology}}\\
 &=\langle\,1\,\rangle_{\mathrm{disk}} +
\langle\,1\,\rangle'_{\mathrm{annulus}}+ \cdots
 +\langle\,{\cal
M}_k\,\rangle_{\mathrm{disk}} + \langle\,{\cal M}_k\,\rangle'_{\mathrm{annulus}}
+ \cdots~,
\label{SM}
\end{aligned}
\end{equation}
where $\langle\,1\,\rangle_{\mathrm{topology}}$ denotes the vacuum amplitudes
and $\langle\,{\cal M}_k\,\rangle_{\mathrm{topology}}$ the amplitudes with
moduli insertions. Since the functional integration over the ADHM moduli ${\cal
M}_k$ is explicitly performed in (\ref{Z1}), to avoid double counting only the
contribution of the ``massive'' string excitations has to be taken into account
in computing the higher order terms of $S({\cal M}_k)$. This is the reason of
the $'$ notation in the annulus contributions, which reminds that only the
``massive'' instantonic string excitations must circulate in the loop.

In the semi-classical approximation, which is typical of the instanton calculus,
it is enough to consider the vacuum amplitudes up to one loop and the moduli
interactions at tree level since, as we will see momentarily,
\begin{equation}
\langle\,1\,\rangle_{\mathrm{disk}} =
\mathcal{O}\big(g_a^{-2}\big)\quad,\quad
\langle\,1\,\rangle'_{\mathrm{annulus}} =
\mathcal{O}\big(g_a^{0}\big)\quad,\quad
\langle\,{\cal M}_k\,\rangle_{\mathrm{disk}} =
\mathcal{O}\big(g_a^{0}\big)~,
\label{orders}
\end{equation}
while $\langle\,{\mathcal M}_k\,\rangle_{\mathrm{annulus}}$ or the higher
topology contributions are of higher order in the Yang-Mills coupling
constant. Thus, in this approximation the $k$-instanton partition function is
\begin{equation}
Z_{k}\,=\, {\cal C}_k ~\ee^{\langle\,1\,\rangle_{\mathrm{disk}} +
\langle\,1\,\rangle'_{\mathrm{annulus}}}
 \int d{\mathcal M}_{k}~\ee^{\langle\,{\mathcal M}_k\,\rangle_{\mathrm{disk}}}
~.
\label{Z33}
\end{equation}
Let us now discuss the various terms of this expression in turn.

The dimensional factor ${\mathcal C}_k$ can be easily determined by
counting the dimensions (measured in units of $\alpha'$) of the
various moduli ${\mathcal M}_{k}$ as given in the previous subsections,
and the result is
\begin{equation}
{\mathcal C}_k = \big({\sqrt {\alpha'}}\big)^{-(2N_a-N_F)k}~.
\label{ck}
\end{equation}
Notice the appearance of the 1-loop coefficient $b_1=(2N_a-N_F)$ of the
$\beta$-function of the $\mathcal{N}=2$ SYM theory%
\footnote{We define the
1-loop $\beta$-function as $\beta(g)=-\,(b_1/16\pi^2)g^3$.}.

The vacuum amplitude at tree level $\langle\,1\,\rangle_{\mathrm{disk}}$ is
nothing but the topological normalization of the a disk whose boundary lies on
the $k$ E$5_a$ branes, which is \cite{Polchinski:1994fq,Billo:2002hm}
\begin{equation}
\langle\,1\,\rangle_{\mathrm{disk}}\,\equiv\, \mathcal{D}_{5_a} \,=\,
-\,\frac{8 \pi^2}{g_a^2}\,k~,
\label{1}
\end{equation}
where $g_a$, given in (\ref{gyma}), is interpreted as the Yang-Mills coupling
constant at the string scale $\sqrt{\alpha'}$. Notice that the
vacuum amplitude (\ref{1}) is also minus the value of the classical
instanton action. Using these results we have
\begin{equation}
{\mathcal C}_k~\ee^{\langle\,1\,\rangle_{\mathrm{disk}}} \,=\,
\Lambda^{(2N_a-N_F)k}~,
\label{Lambda}
\end{equation}
where $\Lambda$ is the renormalization group invariant scale of the
$\mathcal{N}=2$ gauge theory on the color branes. On the other hand these
factors do not seem to have an obvious interpretation in terms of the
four-dimensional field theory living on the flavor branes for which the E$5_a$
branes would represent ``exotic'' instantons of truly stringy nature.

The 1-loop vacuum amplitude
\begin{equation}
\langle\,1\,\rangle_{\mathrm{annulus}} \equiv \mathcal{A}_{5_a}
\label{annulusampl}
\end{equation}
is also contributing to the overall normalization factor of the partition
function through its ``primed'' part. We will give its explicit expression in
the next section, where we will also discuss its meaning and relevance for the
instanton calculus.

The last object appearing in $Z_k$ is the tree-level moduli interaction term
$\langle\,{\cal M}_k\,\rangle_{\mathrm{disk}}$ which can be computed following
the procedure explained in Ref. \cite{Billo:2002hm} from the disk scattering
amplitudes among all ADHM moduli in the limit $\alpha'\to 0$ (with $g_a$ fixed).
The result is \cite{Dorey:2002ik,Billo:2006jm}
\begin{eqnarray}
\langle\,{\cal M}_k\,\rangle_{\mathrm{disk}} &=&{\rm tr}_k  \Big\{
2\,\big[\chi^{\dagger},a_\mu\big]\big[\chi,{a}^\mu\big] - \chi^{\dagger}{\bar
w}_{\dot\alpha} w^{\dot\alpha}\chi - \chi{\bar w}_{\dot\alpha}
w^{\dot\alpha} \chi^{\dagger}
\nonumber
\\
&~-&{\rm i}\, \frac{\sqrt 2}{2}\,{\bar \mu}^A \epsilon_{AB}
\mu^B\chi^{\dagger} +{\rm i}\, \frac{\sqrt 2}{4}\,M^{\alpha
A}\epsilon_{AB}\big[\chi^{\dagger},M_{\alpha}^{B}\big] - {\rm i}\,
\frac{\sqrt 2}{2}\sum_{f=1}^{N_F}\,{{\bar\mu}'}{}^f\,\mu'_f \,\chi
\label{sm}
\\
&~+&\ii D_c\Big({\bar
w}_{\dot\alpha}(\tau^c)^{\dot\alpha}_{~\dot\beta}w^{\dot\beta}
+\ii \bar\eta_{\mu\nu}^c
\big[{a}^\mu,{a}^\nu\big]\Big) - \ii
{\lambda}^{\dot\alpha}_{\,A}\Big(\bar{\mu}^A{w}_{\dot\alpha}+
\bar{w}_{\dot\alpha}{\mu}^A  +
\big[a_\mu,{M}^{\alpha A}\big]\sigma^\mu_{\alpha\dot\alpha}\Big)\Big\}~,
\nonumber
\end{eqnarray}
where we have explicitly indicated the sum over the flavor indices and
understood the one on the color indices. Notice that the moduli $D_c$ and
${\lambda}^{\dot\alpha}_{\,A}$ appear only linearly in the last two terms of
(\ref{sm}) and thus act as Lagrange multipliers for the bosonic and fermionic
constraints of the ADHM construction. Notice also that $\langle\,{\cal
M}_k\,\rangle_{\mathrm{disk}}$ is indeed of $\mathcal{O}\big(g_a^{0}\big)$ as
anticipated above, and that it does not depend on the instanton center
$x_0^{\mu}$ nor on its super-partners $\theta^{\alpha A}$ defined in
(\ref{xtheta}). For this reason it is convenient to separate $x_0^{\mu}$ and
$\theta^{\alpha A}$ from the remaining centered moduli, denoted by
$\widehat{\mathcal{M}}_k$, and simplify the notation by setting $\langle\,{\cal
M}_k\,\rangle_{\mathrm{disk}}\equiv
-\,S_{\mathrm{mod}}(\widehat{\mathcal{M}}_k)$. In this way we have
\begin{equation}
Z_{k}= \int d^4x_0\, d^4\theta \,{\widehat Z_{k}}~,
\label{z}
\end{equation}
where
\begin{equation}
{\widehat Z_{k}}= \Lambda^{(2N_a-N_F)k}~ {\rm e}^{\mathcal{A}_{5_a}'}
~\int d{\widehat{\mathcal M}_{k}}\, {\rm e}^{-S_{\rm mod}({\widehat{\cal
M}_{k}})}
\label{zk3}
\end{equation}
is the centered $k$-instanton partition function.

\subsection{Instanton induced prepotential and effective action}
\label{subsec:2}
Let us now briefly discuss how instanton contributions to gauge field
correlation functions are computed in this string set-up.

\begin{figure}
\begin{center}
\begin{picture}(0,0)%
\includegraphics{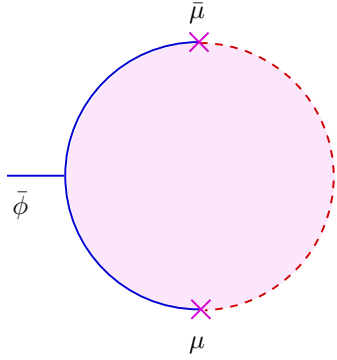}%
\end{picture}%
\setlength{\unitlength}{1579sp}%
\begingroup\makeatletter\ifx\SetFigFont\undefined%
\gdef\SetFigFont#1#2#3#4#5{%
  \reset@font\fontsize{#1}{#2pt}%
  \fontfamily{#3}\fontseries{#4}\fontshape{#5}%
  \selectfont}%
\fi\endgroup%
\begin{picture}(5187,5521)(253,-4955)
\put(3151,314){\makebox(0,0)[lb]{\smash{{\SetFigFont{10}{12.0}{\familydefault}{\mddefault}{\updefault}$\bar\mu$}}}}
\put(376,-2761){\makebox(0,0)[lb]{\smash{{\SetFigFont{10}{12.0}{\familydefault}{\mddefault}{\updefault}$\bar\phi$}}}}
\put(3151,-4861){\makebox(0,0)[lb]{\smash{{\SetFigFont{10}{12.0}{\familydefault}{\mddefault}{\updefault}$\mu$}}}}
\end{picture}%
\end{center}
\caption{The mixed disk representing the coupling of the adjoint scalar field
$\bar\phi$ to the instanton fermionic moduli $\bar\mu$ and $\mu$.}
\label{fig:1}
\end{figure}

The first step is to generalize the moduli action $S_{\rm
mod}({\widehat{\mathcal M}_{k}})$ to include the interactions with gauge fields.
Here for definiteness we will consider only the Coulomb branch of the
$\mathcal{N}=2$ theory, {\it i.e.} we will discuss the interactions with the
adjoint scalar fields. In our semi-classical approximation this is achieved by
computing all possible disk amplitudes with insertions of vertex operators for
instanton moduli and scalar fields as well, like the one represented in Fig.
\ref{fig:1}.

The disk amplitudes that involve the adjoint scalar $\phi$ (or
its conjugate $\bar\phi$) and survive in the limit $\alpha'\to
0$ give rise to the following action \cite{Dorey:2002ik}
\begin{eqnarray}
\label{saf}
S_{\rm mod}(\phi,\bar\phi,m;{\mathcal{M}}_{k} ) & = &-{\rm
tr}_k\Big\{ 2\,[\chi^{\dagger},a'_\mu] [\chi,{a'}^\mu]-
\big(\chi^{\dagger}{\bar w}_{\dot\alpha}-{\bar w}_{\dot\alpha}\,\bar
\phi\big) \big( w^{\dot\alpha}\chi- \phi\,w^{\dot\alpha}\big)
\nonumber\\
&&-\big(\chi{\bar w}_{\dot\alpha} -{\bar w}_{\dot\alpha}\,\phi\big)
\big(w^{\dot\alpha}\chi^{\dagger} -{\bar \phi}\,w^{\dot\alpha}\big)
-{\rm i} \frac{\sqrt 2}{2}\, {\bar \mu}^A \epsilon_{AB}
\big(\mu^B\chi^{\dagger} +\bar \phi\,\mu^B\big)
\\&&+{\rm i}\frac{\sqrt 2}{4}\,M^{\alpha
A}\epsilon_{AB}[\chi^{\dagger},M_{\alpha}^{B}]
-{\rm i} \frac{\sqrt 2}{2}\,\sum_{f=1}^{N_F} \,{{\bar\mu}'}{}^f \big(\mu'_f\chi
+m_f\,\mu'\big)\,
+ S_{\mathrm{constr}}
\Big\}~,
\nonumber
\end{eqnarray}
where $S_{\mathrm{constr}}$ denotes the ADHM constraint part ({\it i.e.} the
last line of (\ref{sm})) which is not modified by gauge fields, and the
hyper-multiplet mass has been taken into account. Notice that $\phi$ and
$\bar\phi$ do not enter into this action on equal footing. For example only
$\bar\phi$, and not $\phi$, couples to the fermionic colored moduli $\bar\mu^A$
and $\mu^B$. This difference has important consequences on the holomorphic
structure of the instanton correlators. Actually there are many other non-zero
disk diagrams with instanton moduli and gauge fields that survive in the field
theory limit. However, as explained in Refs. \cite{Green:2000ke,Billo:2002hm},
the corresponding couplings can be easily obtained from those appearing in the
action (\ref{saf}) by means of supersymmetry Ward identities. In the end, to get
the complete expression one has simply to replace all occurrences of the adjoint
scalars $\phi$ and $\bar\phi$ with the corresponding $\mathcal{N}=2$ chiral and
anti-chiral superfields, $\widehat\phi$ and $\widehat{\overline\phi}$. With this
understanding, the action (\ref{saf}) is then the full result on the Coulomb
branch. All other string amplitudes containing more insertions of moduli or
gauge field vertices or defined on world-sheets of higher topology, either
vanish in the field theory limit or do not contribute in the semi-classical
approximation being of higher order in $g_a$.

The second step is to perform the integration over all moduli of the
action (\ref{saf}) to obtain the $k$-instanton induced gauge effective action
\begin{equation}
\label{effac}
 S_{k} \,=\,\Lambda^{(2N_a-N_F)k}~
 {\rm e}^{\mathcal{A}_{5_a}'}
~\int d^4x_0 \, d^4\theta\,d{\widehat{\mathcal M}_{k}}\, {\rm e}^{-S_{\rm mod}(
\widehat\phi,\widehat{\overline\phi},m;{\widehat{\cal M}_{k}})}~.
\end{equation}
A few comments are in order. First of all, even if $S_{\rm
mod}(\widehat\phi,\widehat{\overline\phi},m;{\widehat{\mathcal{M}}_{k}} )$ has
an explicit dependence on the anti-chiral superfield
${\widehat{\overline\phi}}$, the resulting effective action $S_{k}$ is a {\it
holomorphic} functional of $\widehat\phi$. Indeed, the
${\widehat{\overline\phi}}$ dependence disappears upon integrating over
$\widehat{\mathcal M}_{k}$ as a consequence of the cohomology properties of the
integration measure on the instanton moduli space
\cite{Hollowood:2002ds,Dorey:2002ik,Billo:2006jm}. However, to fully specify the
holomorphic properties of the instanton induced effective action we have to
consider also the contribution of the annulus amplitude that appears in the
prefactor of (\ref{effac}). In principle this term can introduce a
non-holomorphic dependence on the complex and K\"ahler structure moduli of the
compactification space. We will discuss in detail this issue in Sect.
\ref{sec:hol_life} after explicitly computing the annulus amplitude for our
orbifold compactification in the next section.

It is also worth pointing out that among the centered moduli $\widehat{\cal
M}_{k}$ there is the singlet part of the anti-chiral fermions
$\lambda_{\dot\alpha A}$ which is associated to the supersymmetries that are
preserved both by the D9 and by the E5 branes. Thus one may naively think that
instantonic branes cannot generate an F-term, {\it i.e.} an integral on half
superspace, due to the presence of the anti-chiral $\lambda_{\dot\alpha A}$'s
among the integration variables. Actually, this is not true since the
$\lambda_{\dot\alpha A}$'s, including its singlet part, do couple to other
instanton moduli (see the last terms in Eq. (\ref{sm})) and their integration
can be explicitly performed yielding the fermionic ADHM constraints on the
moduli space. Things would be very different instead, if there were no D9$_{a}$
branes, that is if we were discussing the case of the exotic instantons. In this
case, due to the different structure of the charged moduli, the singlet part of
the $\lambda_{\dot\alpha A}$'s would not couple to anything and, unless it is
removed from the spectrum, for example with an orientifold projection
\cite{Argurio:2007vq,Bianchi:2007wy,Ibanez:2007rs}, an integral like the one in
(\ref{effac}) would vanish. In the case of ordinary gauge instantons, instead,
we can write
\begin{equation}
S_{k}\,=\,\int d^4x_0 \, d^4\theta\,\mathcal{F}_{k}({\widehat\phi},m)~,
\label{prep0}
\end{equation}
where the prepotential
\begin{equation}
\label{prep} \mathcal{F}_{k}({\widehat\phi},m)\,=\,\Lambda^{(2N_a-N_F)k}
~ {\rm e}^{\mathcal{A}_{5_a}'}
~\int d{\widehat{\mathcal M}_{k}}\, {\rm e}^{-S_{\rm mod}(
\widehat\phi,\widehat{\overline\phi},m;{\widehat{\mathcal M}_{k}})}
\end{equation}
is the centered instanton partition function in the presence of
$\widehat\phi$. The integral over $\widehat{\mathcal M}_{k}$ can be
performed using localization techniques \cite{Nekrasov:2002qd}.
Choosing a low-energy profile for the adjoint superfield of the form
\begin{equation}
\widehat\phi_{uv} = \widehat\phi_u\,\delta_{uv}
\label{vev}
\end{equation}
where $u,v=1,...,N_a$ and $\sum_u \widehat\phi_u=0$, so that in the effective
theory the gauge group $\mathrm{SU}(N_a)$ is generically broken to
$\mathrm{U}(1)^{N_a-1}$, the prepotential for $k=1$ turns out to be
\begin{equation}
\mathcal{F}_{1}({\widehat\phi},m)= \Lambda^{2N_a-N_F}
~ {\rm e}^{\mathcal{A}_{5_a}'}
~\sum_{u=1}^{N_a}\Bigg[
\prod_{v\not=u}\frac{1}{(\widehat\phi_v-\widehat\phi_u)^2}\,\prod_{f=1}^{N_F}
(\widehat\phi_u+m_f)\Bigg]~.
\label{prep1}
\end{equation}
Similar closed form expressions can be obtained also for higher values of $k$
(see {\it e.g.} Ref. \cite{Dorey:2002ik}). However, for our future
considerations the only relevant feature is that the prepotential
$\mathcal{F}_{k}({\widehat\phi},m)$ is a homogeneous function of its variables,
and specifically
\begin{equation}
\mathcal{F}_{k}(\xi\,{\widehat\phi},\xi\,m) =
\xi^{2-(2N_a-N_F)k}\,\mathcal{F}_{k}({\widehat\phi},m)
\label{homo}
\end{equation}
as one can check from the definition (\ref{prep}).

It is also convenient to write the effective action $S_k$ in terms of (abelian)
$\mathcal{N}=1$ superfields, by decomposing the $\mathcal{N}=2$ superfield
$\widehat\phi$ into its $\mathcal{N}=1$ components $\phi$ and $W_{\alpha}$. Then
we have
\begin{equation}
\begin{aligned}
S_k = \Lambda^{(2N_a-N_F)k}&
~ {\rm e}^{\mathcal{A}_{5_a}'}
\Bigg\{\int d^4x_0 \,d^2\theta\, \Big[\frac1{2g_a^2}\,\tau_{uv}(\phi,m)
\,W^\alpha_u
W_{\alpha v}\Big]
\\
+\int& d^4x_0\, d^2\theta\,d^2\bar\theta
\,\Big[\frac{1}{g_a^2}\,\bar \phi_u \Phi_u^{\mathrm{D}}(\phi,m)\Big]\Bigg\}~,
\end{aligned}
\label{sk1}
\end{equation}
where the functions $\tau$ and $\Phi^{\mathrm{D}}$ are defined by
\begin{equation}
g_a^2\,\left(\frac{\partial^2
\mathcal{F}_{k}}{\partial{\widehat\phi}_u\partial{\widehat\phi}_v}
\right)_{\widehat\phi=\phi}=
\Lambda^{(2N_a-N_F)k}
~ {\rm
e}^{\mathcal{A}_{5_a}'}\,\tau_{uv}(\phi,m)
\label{tau}
\end{equation}
and
\begin{equation}
g_a^2\,\left(\frac{\partial
\mathcal{F}_{k}}{\partial{\widehat\phi}_u}\right)_{\widehat\phi=\phi}=
\Lambda^{(2N_a-N_F)k}
~ {\rm
e}^{\mathcal{A}_{5_a}'}\,\Phi_u^{\mathrm{D}}(\phi,m)~.
\label{phidual}
\end{equation}
All these expressions are written in terms of the canonically
normalized fields, but using the homogeneous property (\ref{homo})
of the prepotential and the rescalings (\ref{phiC}) and (\ref{mf}),
it is straightforward to translate the above result in the supergravity
basis, getting
\begin{eqnarray}
S_k = \Lambda^{(2N_a-N_F)k}\,{\rm
e}^{\mathcal{A}_{5_a}'}&&(g_a\sqrt{K_\Phi})^{(N_F-2N_a)k}\,
\Bigg\{\int d^4x_0 \,d^2\theta\, \Big[\frac1{2g_a^2}\,\tau_{uv}(\Phi,M)
\,W^\alpha_u
W_{\alpha v}\Big]
\nonumber\\
+&&\int d^4x_0\, d^2\theta\,d^2\bar\theta
\,\Big[K_\Phi\,\bar \Phi_u \Phi_u^{\mathrm{D}}(\Phi,M)\Big]\Bigg\}~.
\label{sk2}
\end{eqnarray}
In the following two sections we will carefully analyze the contribution of the
annulus amplitude to the prefactor of the effective action $S_k$ and discuss its
relevance for the holomorphicity properties of the final expression.

\section{The r\^{o}le of annulus amplitudes}
\label{sec:mix_ann}
We now consider in detail the amplitude $\mathcal{A}_{5_a}$ whose ``primed''
part appears in the prefactor of the non-perturbative effective action. This
annulus amplitude represents the 1-loop vacuum energy due to the open strings
with at least one end point on the wrapped instantonic branes. Because of
supersymmetry, the annulus amplitude associated to the E$5_a$/E$5_a$ strings
identically vanishes, so that $\mathcal{A}_{5_a}$ receives contributions only
from mixed annuli with one boundary on the E$5_a$'s and the other on the D$9$
branes. These mixed amplitudes describe the 1-loop contributions of the charged
instantonic open strings ({\it i.e.} the E$5_a$/D$9_a$ and D$9_a$/E$5_a$
strings) and of the flavored instantonic open strings ({\it i.e.} the
E$5_a$/D$9_b$ and D$9_b$/E$5_a$ strings). Their explicit expressions will be
determined in Sect. \ref{subsec:expl_comp}, but before doing this  we present in
the next subsection a general argument that explains their meaning and their
relation with the running gauge coupling constant.

\subsection{The mixed annuli and the running gauge coupling constant}
\label{subsec:margc}
Let us consider the gauge kinetic term at tree level
\begin{equation}
S = \frac{1}{g^2}\,\int d^4x ~{\rm
Tr}\,\Big\{\frac{1}{2}\,F_{\mu\nu}^2\Big\}~.
\label{gautree}
\end{equation}
If we take a constant magnetic field whose only
non-zero component is $F_{23} = f T$  where $T$ is a specific generator
of the gauge group, then the action (\ref{gautree}) simply becomes
\begin{equation}
S(f)  = \frac{V_4 \,f^2}{2\,g^2}~,
\label{gautreea}
\end{equation}
where $V_4$ is the (regularized) volume of space-time.
On the other hand, if we consider an instanton configuration with
charge $k$, then the classical action (\ref{gautree}) is
\begin{equation}
S_{\mathrm{inst}}  =  \frac{8 \pi^2 k}{g^2}~.
\label{gautreeb}
\end{equation}
{F}rom these formulas it is immediate to realize that
\begin{equation}
\frac{S_{\mathrm{inst}}}{8 \pi^2 k} = \frac{S(f)''}{V_4}~,
\label{relation}
\end{equation}
where $''$ means second derivative with respect to $f$. Such a
relation simply expresses the equality of the gauge coupling
constant computed in two different backgrounds.

In the case of supersymmetric theories the same relation (\ref{relation}) holds
also at the quantum level, after taking into account the 1-loop corrections. In
fact, in the constant $f$ background the action (\ref{gautreea}) gets replaced
by
\begin{equation}
S(f) + S^{\mathrm{1-loop}}(f)  = \frac{V_4 \,f^2}{2\,g^2(\mu)}~,
\label{gauloopa}
\end{equation}
where $g(\mu)$ is the running coupling constant at scale $\mu$,
{\it i.e.}
\begin{equation}
\frac{1}{g^2(\mu)}= \frac{1}{ g^2 }
+ \frac{b_1}{1 6 \pi^2} \log \frac{\mu^2}{\Lambda_{\mathrm{UV}}^2}
\label{running0}
\end{equation}
with $\Lambda_{\mathrm{UV}}$ being the ultra-violet cutoff and $b_1$ the 1-loop
coefficient of the $\beta$-function. Similarly, if we consider 1-loop
fluctuations around the instanton background, the action (\ref{gautreeb}) is
simply replaced by
\begin{equation}
S_{\mathrm{inst}} + S_{\mathrm{inst}}^{\mathrm{1-loop}} =
\frac{8 \pi^2 k}{g^2 (\mu)}~.
\label{gauloopb}
\end{equation}
Indeed, in a supersymmetric theory the 1-loop determinants of the non-zero-modes
fluctuations around the instanton cancel out \cite{D'Adda:1977ur} and the only
effect is the renormalization of the gauge coupling constant. Comparing
(\ref{gauloopa}) and (\ref{gauloopb}) we easily see that the same relation
(\ref{relation}) holds also for the 1-loop corrected actions.

\begin{figure}
\begin{minipage}[t]{0.45\linewidth}
\centering
{
\begin{picture}(0,0)%
\includegraphics{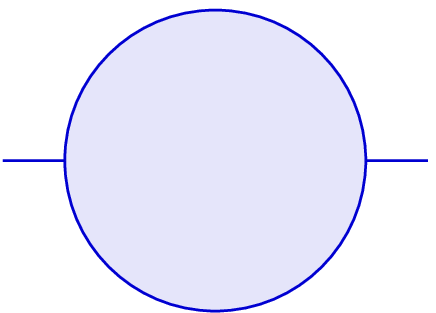}%
\end{picture}%
\setlength{\unitlength}{1579sp}%
\begingroup\makeatletter\ifx\SetFigFont\undefined%
\gdef\SetFigFont#1#2#3#4#5{%
  \reset@font\fontsize{#1}{#2pt}%
  \fontfamily{#3}\fontseries{#4}\fontshape{#5}%
  \selectfont}%
\fi\endgroup%
\begin{picture}(5166,4105)(418,-3989)
\put(526,-1861){\makebox(0,0)[lb]{\smash{{\SetFigFont{10}{12.0}{\familydefault}{\mddefault}{\updefault}$f$}}}}
\put(5026,-1861){\makebox(0,0)[lb]{\smash{{\SetFigFont{10}{12.0}{\familydefault}{\mddefault}{\updefault}$f$}}}}
\put(2926,-136){\makebox(0,0)[lb]{\smash{{\SetFigFont{10}{12.0}{\familydefault}{\mddefault}{\updefault}$9_a$}}}}
\put(526,-1861){\makebox(0,0)[lb]{\smash{{\SetFigFont{10}{12.0}{\familydefault}{\mddefault}{\updefault}$f$}}}}
\end{picture}%
}
\caption{The amplitude $\mathcal{D}_{9_a}(f)$: a disk whose boundary lies on the
D$9_a$ branes, with the insertion of the gauge field at the quadratic order.}
\label{fig:2a}
\end{minipage}
\hskip 0.3cm
\begin{minipage}[t]{0.45\linewidth}
\centering
{
\begin{picture}(0,0)%
\includegraphics{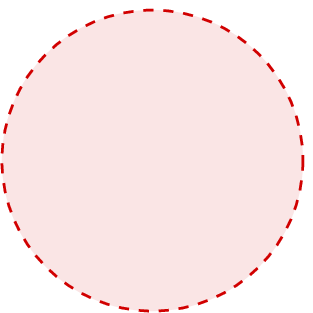}%
\end{picture}%
\setlength{\unitlength}{1579sp}%
\begingroup\makeatletter\ifx\SetFigFont\undefined%
\gdef\SetFigFont#1#2#3#4#5{%
  \reset@font\fontsize{#1}{#2pt}%
  \fontfamily{#3}\fontseries{#4}\fontshape{#5}%
  \selectfont}%
\fi\endgroup%
\begin{picture}(3658,4105)(1172,-3989)
\put(2926,-136){\makebox(0,0)[lb]{\smash{{\SetFigFont{10}{12.0}{\familydefault}{\mddefault}{\updefault}$5_a$}}}}
\end{picture}%
}
\caption{The amplitude $\mathcal{D}_{5_a}$: a disk whose boundary lies on $k$
wrapped E$5_a$ branes.}
\label{fig:2b}
\end{minipage}
\end{figure}

We now show how to rephrase the previous arguments in string theory. As
explained in Sect. \ref{sec:n2_mod}, to obtain $S(f)$ at tree-level we can take
a stack of D$9_a$ branes wrapped on a six-torus and compute the DBI action
(\ref{dbi}) in a constant background gauge field, choosing as before $F_{23}=fT$
and then expanding it to quadratic order in $f$. The result is precisely Eq.
(\ref{gautreea}) with the coupling constant $g_a$ given in (\ref{gyma}). This is
equivalent to compute a tree-level amplitude $\mathcal{D}_{9_a}(f)$ described by
a disk with two insertions of vertex operators for $f$ along its boundary which
lies on the D$9_a$ branes (see Fig. \ref{fig:2a}). More precisely, in Euclidean
signature such amplitude is minus the action $S(f)$, namely
\begin{equation}
\mathcal{D}_{9_a}(f)= -\frac{V_4 \,f^2}{2\,g^2_a}~.
\label{disk9}
\end{equation}

On the other hand, in our string model the classical instanton action
$S_{\mathrm{inst}}$ is obtained from the vacuum amplitude $\mathcal{D}_{5_a}$ on
a disk whose boundary lies on $k$ wrapped E$5_a$ branes as we already explained
in Eq. (\ref{1}), graphically represented in Fig. \ref{fig:2b}, which we rewrite
here for convenience:
\begin{equation}
\mathcal{D}_{5_a} = -\frac{8\pi^2 k}{g^2_a}~.
\label{disk5}
\end{equation}
Thus, from (\ref{disk9}) and (\ref{disk5}) we straightforwardly obtain a
relation between the vacuum disk amplitude with E$5_a$ boundary
conditions and the 2-point function on a disk with D$9_a$ boundary
conditions, namely
\begin{equation}
\frac{\mathcal{D}_{5_a}}{8\pi^2 k} = \frac{\mathcal{D}_{9_a}(f)''}{V_4}
\label{relation1}
\end{equation}
in strict analogy with the field theory result (\ref{relation}).

\begin{figure}
\begin{minipage}[t]{0.45\linewidth}
\centering
{
\begin{picture}(0,0)%
\includegraphics{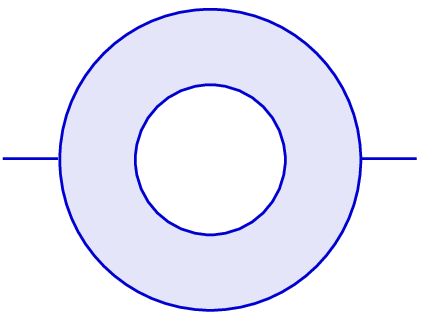}%
\end{picture}%
\setlength{\unitlength}{1579sp}%
\begingroup\makeatletter\ifx\SetFigFont\undefined%
\gdef\SetFigFont#1#2#3#4#5{%
  \reset@font\fontsize{#1}{#2pt}%
  \fontfamily{#3}\fontseries{#4}\fontshape{#5}%
  \selectfont}%
\fi\endgroup%
\begin{picture}(5054,4120)(493,-4004)
\put(541,-1831){\makebox(0,0)[lb]{\smash{{\SetFigFont{10}{12.0}{\familydefault}{\mddefault}{\updefault}$f$}}}}
\put(5071,-1846){\makebox(0,0)[lb]{\smash{{\SetFigFont{10}{12.0}{\familydefault}{\mddefault}{\updefault}$f$}}}}
\put(2926,-136){\makebox(0,0)[lb]{\smash{{\SetFigFont{10}{12.0}{\familydefault}{\mddefault}{\updefault}$9_a$}}}}
\end{picture}%
}
\caption{The amplitudes $\mathcal{A}_{9_a;9_a}(f)$ or
$\mathcal{A}_{9_a;9_b}(f)$ correspond to annuli where one boundary lies on the
D$9_a$ branes and carries two insertions of $f$, while the other boundary lies,
respectively, on a D$9_a$ or a D$9_b$ brane.}
\label{fig:3a}
\end{minipage}
\hskip 0.3cm
\begin{minipage}[t]{0.45\linewidth}
\centering
{
\begin{picture}(0,0)%
\includegraphics{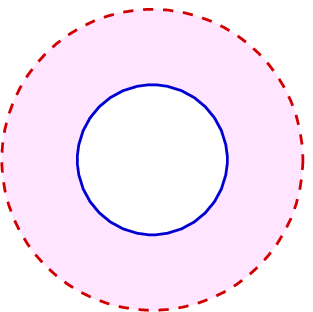}%
\end{picture}%
\setlength{\unitlength}{1579sp}%
\begingroup\makeatletter\ifx\SetFigFont\undefined%
\gdef\SetFigFont#1#2#3#4#5{%
  \reset@font\fontsize{#1}{#2pt}%
  \fontfamily{#3}\fontseries{#4}\fontshape{#5}%
  \selectfont}%
\fi\endgroup%
\begin{picture}(3658,4120)(1187,-4004)
\put(2926,-136){\makebox(0,0)[lb]{\smash{{\SetFigFont{10}{12.0}{\familydefault}{\mddefault}{\updefault}$5_a$}}}}
\end{picture}%
}
\caption{The amplitudes $\mathcal{A}_{5_a;9_a}$ or $\mathcal{A}_{5_a;9_b}$
correspond to annuli where one boundary lies on the E5$a$ branes, while the
other boundary lies, respectively, on a D$9_a$ or a D$9_b$ brane.}
\label{fig:3b}
\end{minipage}
\end{figure}

The same kind of relation holds also for 1-loop amplitudes. In fact, in the
constant gauge field background the 1-loop correction to the classical action is
obtained by computing the vacuum amplitude on an annulus%
\footnote{In orientifold models also the M\"obius strip has to be considered.}
with one boundary on the brane with $f$ and the second boundary on the other
branes~\cite{Di Vecchia:2003ae,Di Vecchia:2005vm},
and then expanding the result to second order in $f$. This is equivalent to
compute the 2-point function $\mathcal{A}_{9_a;9_a}(f)$ represented in Fig.
\ref{fig:3a} where the loop is spanned by the D$9_a$/D$9_a$ strings. If also
flavor branes are present, we should consider also the annulus amplitude
$\mathcal{A}_{9_a;9_b}(f)$ with D$9_a$/D$9_b$ and D$9_b$/D$9_a$ strings
circulating in the loop. These open string amplitudes exhibit both UV and IR
divergences. The UV divergences, corresponding to IR divergences in the dual
closed string channel, cancel in consistent tadpole-free models; even if in this
paper we take only a local point of view, we assume that globally the closed
string tadpoles are absent so that we can ignore the UV divergences. On the
other hand, introducing a cutoff $\mu$ to regulate the IR divergences, the above
annulus amplitudes take the following form \cite{Lust:2003ky,Di Vecchia:2005vm}
\begin{equation}
\label{s1log}
\mathcal{A}_{9_a}(f)\,\equiv\,\mathcal{A}_{9_a;9_a}(f)+\mathcal{A}_{9_a;9_b}(f)
\,=\,-\, \frac{V_4\, f^2}{2}\left(\frac{b_1}{16 \pi^2} \log(\alpha'\mu^2) +
\Delta_a\right)~.
\end{equation}
The logarithmic term accounts for the massless open string states circulating in
the loop and is thus proportional to the coefficient $b_1$ of the
$\beta$-function. The finite term $\Delta_a$ originates from the integration
over massive states and represents the threshold corrections. In principle,
these are due to excited string states and/or to Kaluza-Klein  modes arising
from  the  compactification of the six extra dimensions. In $\mathcal{N}=2$
models, however, only the Kaluza-Klein partners of the massless string states do
contribute while the excited string states cancel each other \cite{Di
Vecchia:2003ae,Di Vecchia:2005vm}. Notice also that in (\ref{s1log}) the r\^ole
of the UV cutoff is played naturally by the string length.

Let us now consider the instanton background. As we mentioned above, the 1-loop
amplitudes in this case correspond to mixed annulus diagrams with one boundary
on the instantonic E$5_a$ branes and the other boundary on the color D$9_a$
branes or on the flavor D$9_b$ branes. These amplitudes, denoted
$\mathcal{A}_{5_a;9_a}$ and $\mathcal{A}_{5_a;9_b}$ respectively, are
represented in Fig. \ref{fig:3b} and will be explicitly computed in the
following subsection. However, even before computing them, we can understand
their meaning using the relation (\ref{relation1}) which allows to trade the
boundary on the E$5_a$ for a boundary on the D$9_a$'s with a constant field $f$.
Thus, in a supersymmetric model the total annulus amplitude $\mathcal{A}_{5_a}$
must account only for the 1-loop correction to the gauge coupling constant in a
$k$ instanton background and, after regulating the IR divergences, we expect to
find
\begin{equation}
\label{s1loginst}
\mathcal{A}_{5_a} \,\equiv\,\mathcal{A}_{5_a;9_a}+\mathcal{A}_{5_a;9_b}
\,=\,-\,
8\pi^2k\left(\frac{b_1}{16 \pi^2} \log(\alpha'\mu^2) + \Delta_a\right)\,
~.
\end{equation}
Notice that in this context the $\beta$-function coefficient $b_1$ arises from
the counting (with appropriate sign and weight) of the bosonic and fermionic
ground states of mixed open strings with one end point on the E$5_a$ branes,
{\it i.e.} from the charged and flavored instanton moduli that we listed in
Sect. \ref{subsec:1}. We will elaborate more on this point in the following
subsection after the explicit computation of the mixed annulus amplitudes.

{F}rom (\ref{s1log}) and (\ref{s1loginst}) it immediately follows that
\begin{equation}
\frac{\mathcal{A}_{5_a}}{8\pi^2 k} = \frac{\mathcal{A}_{9_a}(f)''}{V_4}~,
\label{relation2}
\end{equation}
which is the natural generalization of (\ref{relation1}) at 1-loop. The relation
(\ref{relation2}) between the annulus with a boundary on the instantonic brane
and the annulus with a constant gauge field $f$, which has been noticed in
Refs.~\cite{Abel:2006yk,Akerblom:2006hx}, is the strict analogue of the field
theory relation (\ref{relation}) and it simply expresses the equality of the
(running) gauge coupling constant computed in two different backgrounds. {F}rom
our arguments it also follows that in supersymmetric models the annulus
amplitudes with wrapped Euclidean branes and no moduli insertions, contrarily to
some claims in the literature, seem not to be related to the 1-loop determinants
of the non-zero mode fluctuations around the instanton background, which in fact
are known to exactly cancel out because of supersymmetry \cite{D'Adda:1977ur}.

\subsection{The explicit form of the annulus amplitude $\mathcal{A}_{5_a}$}
\label{subsec:expl_comp}
The annulus amplitude $\mathcal{A}_{5_a}$ is the 1-loop free energy of the open
strings suspended between the E$5_a$ branes and the D9 branes and, as indicated
in (\ref{s1loginst}), consists of a contribution from the charged instanton
sector, $\mathcal{A}_{5_a;9_a}$, and a contribution from the flavored instanton
sector, $\mathcal{A}_{5_a;9_b}$. In turn each of these individual contributions
is a sum of two terms corresponding to the two possible orientations of the open
strings, {\it e.g.}
\begin{equation}
\mathcal{A}_{5_a;9_a} \equiv \mathcal{A}(9_a/5_a) + \mathcal{A}(5_a/9_a)~,
\label{ampli5a9a}
\end{equation}
and similarly for the flavored strings. Let us now give some
details on these amplitudes, starting from the charged sector.

\paragraph{The charged instanton sector}
For a given open string orientation, the annulus amplitude in the charged sector
has the following schematic form
\begin{equation}
\label{ma0}
\mathcal{A}(9_a/5_a) = \int _0^\infty \frac{d\tau}{2\tau}\Big[
\Tr_{\mathrm{NS}} \left(P_{\mathrm{GSO}}^{(9_a/5_a)} \,P_{\mathrm{orb}}\,
q^{L_0}\right)
- \Tr_{\mathrm{R}}\left(P_{\mathrm{GSO}}^{(9_a/5_a)}\,P_{\mathrm{orb}}\,
q^{L_0}\right)\Big]~,
\end{equation}
where
\begin{equation}
\label{gso9a5a}
P_{\mathrm{GSO}}^{(9_a/5_a)} = \frac{1 + (-1)^F}{2}
\end{equation}
is the GSO projector, $P_{\mathrm{orb}}$ is the orbifold projector (see Eq.
(\ref{projorb})) and $q= \exp(-2\pi\tau)$. The amplitude $\mathcal{A}(5_a/9_a) $
corresponding to open strings with opposite orientation is analogous to
(\ref{ma0}), but we must consider the possibility that the GSO projection
$P_{\mathrm{GSO}}^{(5_a/9_a)}$ to be employed in this case may be different from
the one in (\ref{gso9a5a}). We will argue that this is indeed the case in the R
sector.

The traces in (\ref{ma0}) are taken over the states in the CFT of the open
strings with D$9_a$/E$5_a$ boundary conditions and over the Chan-Paton indices
as well. The CFT contains various components: the string fields $X^\mu$ and
$\psi^\mu$ in the space-time directions, those along the orbifold
$(\mathcal{T}_2^{(1)}\times\mathcal{T}_2^{(2)})/\mathbb{Z}_2$, those along
$\mathcal{T}_2^{(3)}$ and the ghost/superghost system. Let us now discuss
briefly the contributions of these various components.

The fields $X^\mu$ and $\psi^\mu$ in the space-time directions have
Neumann-Dirichlet conditions, as discussed in Sect. \ref{sec:inst_calc},
and hence are twisted by $1/2$; in particular, their contribution to the trace
in the NS$(-1)^F$ structure vanishes because of the fermionic zero-modes.

Moving to the internal directions, all fields $Z^i$ and $\Psi^i$ have
Neumann-Neumann boundary conditions and thus are untwisted, but for $i=1,2$ they
are reflected by the $\mathbb{Z}_2$ action so that they yield different non-zero
mode contributions depending on whether the orbifold generator $h$ is inserted
or not in the trace. On the other hand, the fields $Z^3$ and $\Psi^3$ are not
acted upon by the orbifold and their non-zero mode contributions cancel exactly
against those from the ghost/superghost system. Concerning the zero-modes, the
trace over the discretized momenta of the bosonic fields $Z^i$ gives a
contribution of the form $\mathcal{Y}^{(1)}\mathcal{Y}^{(2)}\mathcal{Y}^{(3)}$
where
\begin{equation}
\label{ma3}
\mathcal{Y}^{(i)}\equiv
\sum_{(r_1 , r_2) \in \mathbb{Z}^2} q^{\frac{r_p}{n_a^{(i)}}\,
\mathcal{G}^{pq}_{(i)}\,\frac{r_q}{n_a^{(i)}}} =
\sum_{(r_1 , r_2) \in \mathbb{Z}^2} q^{
\frac{ T_{2}^{(i)}}{U_{2}^{(i)}}
\frac{ | r_{1} U^{(i)}-r_{2} |^2}{ |n_{a}^{(i)}T^{(i)} - m_{a}^{(i)}|^2}}
=
\sum_{(r_1 , r_2) \in \mathbb{Z}^2} q^{
\frac{ | r_{1} U^{(i)}-r_{2} |^2}{U_{2}^{(i)}T_{2}^{(i)}|\ell_a^{(i)}|^2}}~.
\end{equation}
In this expression $\mathcal{G}^{pq}_{(i)}$ is the
inverse open string metric on the $i$-th torus%
\footnote{The open string is defined as $
\mathcal{G}_{(i)} = \left(G_{(i)} + B_{(i)} - 2 \pi\alpha' F_{(i)}\right)
\,G_{(i)}^{-1}\,
\left(G_{(i)} - B_{(i)} + 2 \pi\alpha' F_{(i)}\right)$.
}
and in the last step we have used the definition (\ref{vai}). Notice however
that when $h$ is inserted in the trace, only the zero-momentum states along the
first two tori survive and thus in this case the bosonic zero-mode contribution
reduces to $\mathcal{Y}^{(3)}$.

Finally, let us consider the fermionic zero modes of the $\Psi^i$ fields in the
R sector. There are eight zero modes corresponding to the following states:
\begin{equation}
\label{mua}
\begin{aligned}
|\Delta S_A\rangle &\equiv
\Big\{|\Delta\,S_{-++}\rangle\,,\,|\Delta\,S_{+-+}\rangle
\,,\,|\Delta\,S_{++-}\rangle\,,\,|\Delta\,S_{---}\rangle\Big\}~,
\\
|\Delta S^A\rangle &\equiv
\Big\{|\Delta\,S^{+--}\rangle\,,\,|\Delta\,S^{-+-}\rangle
\,,\,|\Delta\,S^{--+}\rangle\,,\,|\Delta\,S^{+++}\rangle\Big\}~,
\end{aligned}
\end{equation}
where $S^A$ and $S_A$ are the spin fields in the six-dimensional
internal space. The action of $(-1)^F$ on these states is defined
to be
\begin{equation}
(-1)^F\,|\Delta S_A\rangle \,=\,+\,|\Delta S_A\rangle \quad,\quad
(-1)^F\,|\Delta S^A\rangle \,=\,-\,|\Delta S^A\rangle~,
\label{f}
\end{equation}
while the orbifold action is given in (\ref{gorb}). Thus the GSO projection
(\ref{gso9a5a}) selects the states $\ket{\Delta S_A}$ that are associated to
four charged fermionic moduli $\mu^A$, of which only two are $h$-invariant and
appear in the physical spectrum of the D$9_a$/E$5_a$ strings, as described in
Sec. \ref{subsec:1}. With this information it is possible to evaluate in a
straightforward manner the contribution of these fermionic zero-modes to the
trace in the 1-loop amplitude. In the odd spin structure, because of the
insertion of $(-1)^F$, this trace vanishes but simultaneously the superghost
zero-modes give a divergent contribution, which makes the entire expression
ill-defined. However, as discussed in Ref.~\cite{Billo:1998vr}, there exists a
suitable regularization procedure for both contributions which makes their
product well-defined and actually finite. In particular, it turns out (see for
example the discussion after Eq. (B.11) of Ref.~\cite{Billo:2000yb}) that the
trace over the fermionic zero-modes vanishes when we insert $(-1)^F$ or $h$,
while it equals $8/2=4$ when there is no insertion or the insertion of
$(-1)^Fh$.

Altogether, collecting all the previous information, we can obtain the explicit
expression for the amplitude $\mathcal{A}(9_a/5_a)$. In the NS spin structure we
find
\begin{equation}
\begin{aligned}
\mathcal{A}(9_a/5_a)_{\mathrm{NS}} &\equiv\,
\frac{1}{2}\int _0^\infty \frac{d\tau}{2\tau}\,
\Tr_{\mathrm{NS}} \left(P_{\mathrm{orb}}\,
q^{L_0}\right)
\\
&=
\frac{N_a k}2\int _0^\infty \frac{d\tau}{2\tau} \Bigg[\frac 12
\Bigg( \frac{\theta_2(0)^2\,\,\theta_3(0)^2}{\theta_4(0)^2\,\theta_1'(0)^2} \,
\mathcal{Y}^{(1)}\mathcal{Y}^{(2)}\mathcal{Y}^{(3)}
+ 4\,
\mathcal{Y}^{(3)}
 \Bigg)\Bigg]~,
 \end{aligned}
\label{aaNS}
\end{equation}
where the $\theta_a$'s are the Jacobi $\theta$-functions (we follow the
conventions of Appendix A of Ref.~\cite{Di Vecchia:2005vm}). The second term in
(\ref{aaNS}) contains the insertion of $h$, upon which the non-zero modes
contributions along the orbifold and the space-time directions cancel each
other. The factor of $1/2$ inside the square bracket comes from the orbifold
projector. As argued above, the NS$(-1)^F$ structure vanishes identically. In
the R sector, taking into account the minus sign due to spin-statistics, we get
\begin{equation}
\begin{aligned}
\mathcal{A}(9_a/5_a)_{\mathrm{R}}
&\equiv
\,-\,\frac{1}{2}\int _0^\infty \frac{d\tau}{2\tau}\,
\Tr_{\mathrm{R}} \left(P_{\mathrm{orb}}\,
q^{L_0}\right)
\\
&
= -\,\frac{N_a k}2 \int _0^\infty \frac{d\tau}{2\tau} \Bigg[\frac 12\Bigg(
\frac{\theta_3(0)^2\,\,\theta_2(0)^2}{\theta_4(0)^2\,\theta_1'(0)^2} \,
\mathcal{Y}^{(1)}\mathcal{Y}^{(2)}\mathcal{Y}^{(3)}\Bigg)\Bigg]~,
\end{aligned}
\label{aaR}
\end{equation}
which comes entirely from the term with no insertion of $h$. Finally, the odd
spin structure R$(-1)^F$ receives a contribution only when $h$ is inserted, and
reads
\begin{equation}
\begin{aligned}
\mathcal{A}(9_a/5_a)_{\mathrm{R}(-1)^F} &\equiv
\,-\,\frac{1}{2}\int _0^\infty \frac{d\tau}{2\tau}\,
\Tr_{\mathrm{R}} \left((-1)^F\,P_{\mathrm{orb}}\,
q^{L_0}\right)
\\
&=-\,
\frac{N_a k}2 \int _0^\infty \frac{d\tau}{2\tau} \Big[\,\frac 12 \,\big( 4
\, \mathcal{Y}^{(3)}\big)\Big]~.
\end{aligned}
\label{aaRmf}
\end{equation}
Then the full GSO-projected amplitude for the D$9_a$/E$5_a$ strings is
\begin{equation}
 \label{fullaa}
\mathcal{A}(9_a/5_a) = \mathcal{A}(9_a/5_a)_{\mathrm{NS}} +
\mathcal{A}(9_a/5_a)_{\mathrm{R}} +
\mathcal{A}(9_a/5_a)_{\mathrm{R}(-1)^F} = 0~,
\end{equation}
where we have inserted the results (\ref{aaNS}), (\ref{aaR}) and
(\ref{aaRmf}).

However, we have to consider also the amplitude $\mathcal{A}(5_a/9_a)$ which is
the 1-loop vacuum energy of open strings with the opposite orientation. The only
subtlety occurs in the R sector. In this case we have again eight fermionic
ground states, namely $\ket{\bar\Delta S_A}$ and  $\ket{\bar\Delta S^A}$ which
differ from the states (\ref{mua}) only because they contain the anti-twist
$\bar\Delta$ in place of $\Delta$. The $(-1)^F$ parity on these states must be
defined consistently with the previous definition (\ref{f}). To do so, let us
observe that
\begin{equation}
\langle \,\bar\Delta S^A\,|\,\Delta S_B\rangle = \delta^A_B~.
\label{scalarproduct}
\end{equation}
This pairing, together with (\ref{f}), implies the following parity assignments
\begin{equation}
(-1)^F\,|\bar\Delta S_A\rangle \,=\,-\,|\bar\Delta S_A\rangle~,\quad
(-1)^F\,|\bar\Delta S^A\rangle \,=\,+\,|\bar\Delta S^A\rangle~.
\label{f1}
\end{equation}
As discussed after Eq. (\ref{vertmu}), the physical spectrum of the $5_a$/$9_a$
strings contains the moduli $\bar\mu^A$ with the {\it same} chirality as the
$\mu^A$. Thus, the GSO projection must select the corresponding states, namely
$|\bar\Delta S_A\rangle$ which are odd under $(-1)^F$. Therefore, in the R
sector of the $5_a$/$9_a$ strings we must take
\begin{equation}
P_{\mathrm{GSO}}^{(5_a/9_a)} = \frac{1-(-1)^F}{2}~.
\label{gso5a9a}
\end{equation}
as opposed to (\ref{gso9a5a}). The full GSO-projected amplitude is then
\begin{equation}
 \label{fullaaoo}
\mathcal{A}(5_a/9_a) = \mathcal{A}(5_a/9_a)_{\mathrm{NS}} +
\mathcal{A}(5_a/9_a)_{\mathrm{R}} -
\mathcal{A}(5_a/9_a)_{\mathrm{R}(-1)^F}
\end{equation}
with a crucial minus sign in the odd spin structure as compared to
(\ref{fullaa}). The individual terms in this expression can be computed as
explained above and turn out to be equal to the corresponding ones for the other
orientation, given in Eqs. (\ref{aaNS}), (\ref{aaR}) and (\ref{aaRmf})
respectively. Now, however, due to the different sign in the R$(-1)^F$ sector,
the amplitude $\mathcal{A}(5_a/9_a)$ is not zero. The fact that the annulus
amplitude is different for the two open string orientations when the odd spin
structure is non zero should not come as a surprise; in fact the same thing has
been noticed in other systems with similar features, most notably in the D0/D8
brane systems or their T-duals \cite{Billo:1998vr}.

{From} the above analysis we conclude that the total amplitude (\ref{ampli5a9a})
is
\begin{equation}
\label{atotfin}
\mathcal{A}_{5_a;9_a} = \mathcal{A}(5_a/9_a) = 2 N_a k \int_0^\infty
\frac{d\tau}{2\tau}
\, \mathcal{Y}^{(3)}~.
\end{equation}
In the end only the zero-modes contribute to this annulus amplitude: they
correspond to the charged instanton moduli listed in Sect. \ref{subsec:1}, and
their Kaluza-Klein partners on the torus $\mathcal{T}_2^{(3)}$ that together
reconstruct the sum in $\mathcal{Y}^{(3)}$.

\paragraph{The flavored instanton sector}
Let us now consider the annulus amplitude produced by the instantonic strings
stretching between the flavor D$9_b$ branes and the E$5_a$'s, namely
\begin{equation}
\mathcal{A}_{5_a;9_b} \equiv \mathcal{A}(9_b/5_a) +
\mathcal{A}(5_a/9_b)~.
\label{ampli5a9b}
\end{equation}
The difference with the charged case considered above resides entirely in the
CFT of the string fields $Z^i$ and $\Psi^i$, with $i=1,2$, along the orbifold.
These fields are all twisted by the same angle $\nu_{ba}^{(1)}=\nu_{ba}^{(2)}$,
which in the following will be simply denoted by $\nu$, and none of them has
zero-modes. We have however to include the factor $I_{ba}$, defined in Eq.
(\ref{iab}), related to the number of Landau levels for these magnetized
directions. Another difference is that the fractional branes of type $a$ and $b$
belong to different irreducible representations of the $\mathbb{Z}_2$ orbifold
group, so that $h$ acts on the Chan-Paton factors of the open strings as a minus
sign, and the twisted NS ground state is $h$-even while the twisted R ground
state is $h$-odd as explained in Sect.~\ref{subsec:mattersector}.

Taking all these facts into account, we can write the various contributions to
the annulus amplitude. Let us start with the D$9_b$/E$5_a$ orientation. In the
NS spin structure we have
\begin{equation}
\begin{aligned}
\mathcal{A}(9_b/5_a)_{\mathrm{NS}} &\equiv\,
\frac{1}{2}\int _0^\infty \frac{d\tau}{2\tau}\,
\Tr_{\mathrm{NS}} \left(P_{\mathrm{orb}}\,
q^{L_0}\right)
\\
&=
-\,\frac{N_b I_{ba} k}2\int _0^\infty \frac{d\tau}{2\tau} \Bigg[\frac 12
\Bigg(
\frac{\theta_2(0)^2\,\,\theta_3(\ii\nu\tau)^2}{\theta_4(0)^2\,
\theta_1(\ii\nu\tau)^2}
+\frac{\theta_2(0)^2\,\theta_4(\ii\nu\tau)^2}{\theta_4(0)^2\,
\theta_2(\ii\nu\tau)^2}\Bigg)
\mathcal{Y}^{(3)}
 \Bigg]~.
 \end{aligned}
\label{AbaNS}
\end{equation}
The NS$(-1)^F$ amplitude vanishes because of the space-time fermion zero modes,
as before. In the R sector we find instead
\begin{equation}
\begin{aligned}
\mathcal{A}(9_b/5_a)_{\mathrm{R}} &\equiv\,-\,
\frac{1}{2}\int _0^\infty \frac{d\tau}{2\tau}\,
\Tr_{\mathrm{R}} \left(P_{\mathrm{orb}}\,
q^{L_0}\right)
\\
&=
\frac{N_b I_{ba} k}2\int _0^\infty \frac{d\tau}{2\tau} \Bigg[\frac 12
\Bigg(
\frac{\theta_3(0)^2\,\theta_2(\ii\nu\tau)^2}{\theta_4(0)^2\,
\theta_1(\ii\nu\tau)^2}
+\frac{\theta_3(0)^2\,\theta_1(\ii\nu\tau)^2}{\theta_4(0)^2\,
\theta_2(\ii\nu\tau)^2}\Bigg)
\mathcal{Y}^{(3)}
 \Bigg]~.
 \end{aligned}
\label{AbaR}
\end{equation}
Finally, the R$(-1)^F$ amplitude, to which both the term without $h$ and the one
with $h$ contribute, is
\begin{equation}
\mathcal{A}(9_b/5_a)_{\mathrm{R}(-1)^F} \equiv\,-\,
\frac{1}{2}\int _0^\infty \frac{d\tau}{2\tau}\,
\Tr_{\mathrm{R}} \left((-1)^F\,P_{\mathrm{orb}}\,
q^{L_0}\right)
\,=\,
\frac{N_b I_{ba} k}2\int _0^\infty \frac{d\tau}{2\tau} \,\mathcal{Y}^{(3)}~.
\label{AbaRf}
\end{equation}
Using the Riemann identities
\begin{equation}
 \label{riemann}
\begin{aligned}
&\theta_2(0)^2\,\theta_3(\ii\nu\tau)^2 -
\theta_3(0)^2\,\theta_2(\ii\nu\tau)^2 =
\theta_4(0)^2\,\theta_1(\ii\nu\tau)^2~,
\\
&\theta_2(0)^2\,\theta_4(\ii\nu\tau)^2 -
\theta_3(0)^2\,\theta_1(\ii\nu\tau)^2 =
\theta_4(0)^2\,\theta_2(\ii\nu\tau)^2~,
\end{aligned}
\end{equation}
one can easily see that the GSO projected amplitude vanishes:
\begin{equation}
\mathcal{A}(9_b/5_a)=\mathcal{A}(9_b/5_a)_{\mathrm{NS}}
+
\mathcal{A}(9_b/5_a)_{\mathrm{R}}+\mathcal{A}(9_b/5_a)_{\mathrm{R}(-1)^F}
=0~.
\label{a9b5atot}
\end{equation}

Let us now consider the amplitude $\mathcal{A}(5_a/9_b)$ corresponding to the
other orientation. Just as in the charged case previously discussed, we must be
careful with the GSO projection in the R sector. The same argument presented
above implies that $P_{\mathrm{GSO}}^{(5a/9b)}$ and $P_{\mathrm{GSO}}^{(9b/5a)}$
must be defined in a different way. Indeed, the physical states of the two types
of strings are described by the vertex operators (\ref{vertmup}) and
(\ref{vertmup1}) which both contain the same spin field $S_-$ in the last
complex direction. Thus, if the $\mu'$'s are even under $(-1)^F$, the
${\bar\mu}'$'s, which do not contain the conjugate spin field, must be odd under
$(-1)^F$. Then, the complete GSO projected amplitude for the E$5_a$/D$9_b$
strings reads
\begin{equation}
\mathcal{A}(5_a/9_b)=\mathcal{A}(5_a/9_b)_{\mathrm{NS}}
+
\mathcal{A}(5_a/9_b)_{\mathrm{R}}-\mathcal{A}(5_a/9_b)_{\mathrm{R}(-1)^F}~.
\label{a5a9btot}
\end{equation}
The individual contributions can be computed explicitly as before; one simply
has to replace $\nu \to (1 -\nu)$, which however has no consequences because of
the properties of the $\theta$-functions, and also change the prefactor to $N_b
I_{ab}$, which is also harmless since $I_{ab} = I_{ba}$, see Eq.~(\ref{iab}).
Thus, the various terms in (\ref{a5a9btot}) are equal to the corresponding ones
for the other orientation, given respectively in Eqs.~(\ref{AbaNS}),
(\ref{AbaR}) and (\ref{AbaRf}). Now, however, due to the minus sign in the odd
spin structure, the amplitude $\mathcal{A}(5_a/9_b)$ is not vanishing.

We thus conclude that the total instantonic amplitude in the flavored sector is
\begin{equation}
\label{atotfinba}
\mathcal{A}_{5_a;9_b} = \mathcal{A}(5_a/9_b) = - N_F\, k \int_0^\infty
\frac{d\tau}{2\tau}
\, \mathcal{Y}^{(3)}~,
\end{equation}
where $N_F$ is the number of flavors defined in (\ref{nfl}).

\paragraph{The total amplitude}
Summing the contributions (\ref{atotfin}) and (\ref{atotfinba}) of the charged
and flavor sectors, we finally have
\begin{equation}
{\mathcal A}_{5_a} =
\left(2 N_a - N_F\right) k_a\,
\int _0^\infty \frac{d\tau}{2\tau}\, \mathcal{Y}^{(3)}~.
\end{equation}
This amplitude is proportional to the 1-loop coefficient $b_1$ of the
$\beta$-function of our $\mathcal{N}=2$ theory, {\it i.e.} $b_1=2N_a-N_F$. It is
interesting to notice that in this context this coefficient arises from the
counting of the charged and flavored zero-modes of the instantonic strings. Let
us consider in more detail this contribution, tracing back the NS and R terms
and keeping them distinct. We have
\begin{equation}
\label{aatot0m}
\int _0^\infty \frac{d\tau}{2\tau} \Big[(4 N_a- 2 N_F)k - (2 N_a -N_F)k\Big] =
\Big(n_{\mathrm{bos}} - \frac 12\,
n_{\mathrm{ferm}}\Big) \int _0^\infty \frac{d\tau}{2\tau}~,
\end{equation}
where
\begin{equation}
n_{\mathrm{bos}}=n_{\mathrm{ferm}}=(4 N_a- 2 N_F)k
\label{number0m}
\end{equation}
is the number of bosonic and fermionic moduli in the charged and flavored
sectors, {\it i.e.} the number of $w$'s and $\bar w$'s and the number of
$\mu$'s, $\bar\mu$'s, $\mu'$'s and ${\bar\mu}'$'s. Notice also that the stringy
origin of the factor of 1/2 in (\ref{aatot0m}) is in the (regularized) trace
over the superghost zero-modes of the R sector \cite{Billo:1998vr}.

To obtain the explicit expression of the annulus amplitude ${\mathcal A}_{5_a}$
we have to compute the integral
\begin{equation}
I \equiv
\int_0^{\infty} \frac{d\tau}{\tau}\, \mathcal{Y}^{(3)} =
\int_0^{\infty} \frac{d\tau}{\tau}
\sum_{(r_1 , r_2) \in \mathbb{Z}^2} \ee^{-2\pi\tau\,{
\frac{ | r_{1} U^{(3)}-r_{2} |^2}{U_{2}^{(3)}T_{2}^{(3)}|\ell_a^{(3)}|^2}}
}~.
 \label{kint}
 \end{equation}
The detailed calculation is performed in Appendix \ref{integral}; here we simply
recall that this integral is divergent both in the UV limit $\tau\to 0$, and in
the IR limit $\tau\to \infty$. The UV divergence can be reinterpreted as an IR
divergence in the dual closed string channel after Poisson resummation. We
assume that such a divergence cancels in fully consistent models which satisfy
the tadpole cancellation condition \cite{Lust:2003ky}. Subtracting this
divergence, the integral $I$ can be evaluated by introducing a mass parameter
$m$ which regularizes the IR singularity in the open string channel, and the
final result is (see Eq. (\ref{ku1}))
\begin{equation}
I =
- \log \left(\alpha^\prime m^2\right)  - \log |\eta(U^{(3)}) |^4
- \log\left( U_2^{(3)} T_2^{(3)} |\ell_{a}^{(3)}|^2\right)~,
\label{ku0}
\end{equation}
where $\eta$ is the Dedekind function. Since only one of the two orientations
contributes to ${\mathcal A}_{5_a}$, it is possible, following Refs. \cite{Di
Vecchia:2003ae,Di Vecchia:2005vm}, to take a complex IR cutoff%
\footnote{Notice that in general one should regulate, for consistency, the
contributions of the two orientations with complex conjugate cutoffs \cite{Di
Vecchia:2005vm}.}
\begin{equation}
m= \mu\,\ee^{\ii\varphi}~,
\label{ircutoff}
\end{equation}
so that the instantonic annulus amplitude becomes
\begin{equation}
{\mathcal A}_{5_a}= -b_1\,k\left(\frac12 \log(\alpha'\mu^2)+
\ii\,\varphi+\log |\eta(U^{(3)}) |^2
+ \frac12\,\log\left( U_2^{(3)} T_2^{(3)}
|\ell_{a}^{(3)}|^2\right)\right)
\end{equation}
and has the expected form (\ref{s1loginst}).

\section{The holomorphic life of the D-brane instantons}
\label{sec:hol_life}
In this section we combine the result we have just obtained for the annulus
amplitude with what we have discussed in Sect. \ref{subsec:2} in order to get
the instanton induced corrections to the low-energy effective action of our
$\mathcal{N}=2$ theory.

To this aim, let us first recall that what enters in the instanton calculus is
{\it not} the complete annulus amplitude $\mathcal{A}_{5_a}$, but rather its
``primed'' part $\mathcal{A}_{5_a}^\prime$. This is obtained from
$\mathcal{A}_{5_a}$ by subtracting the logarithmically divergent contribution of
the zero-modes to avoid double counting since the integral over them is
separately performed in an explicit way
\cite{Blumenhagen:2006xt,Akerblom:2006hx}. However, as remarked already in
Refs.~\cite{Kaplunovsky:1994fg,Louis:1996ya}, the UV cutoff that one uses in the
field theory analysis of a string model is the four-dimensional Planck mass
$M_P$, which is related to $\alpha'$ in the following way:
\begin{equation}
\label{mp}
M_P^2\,=\,\frac{1}{\alpha'}\,{\rm e}^{-\phi_{10}}\,s_2~,
\end{equation}
where $\phi_{10}$ is the ten-dimensional dilaton. This means that what we have
to subtract from ${\mathcal A}_{5_a}$ in order to remove the field theory zero
modes contribution is not exactly the $\log(\alpha'\mu^2)$ term. Rather, we have
to write
\begin{equation}
\label{s1loginst3}
\mathcal{A}_{5_a}
\,=\,-8\pi^2k\left(\frac{b_1}{16 \pi^2} \log \frac{\mu^2}{M_P^2}\,+\,
{\widetilde \Delta}_a\right)\,
\end{equation}
with
\begin{equation}
\label{del}
{\widetilde\Delta}_{a} \,=\,
\frac{b_1}{8 \pi^2} \left( \ii\,\varphi+\log |\eta(U^{(3)})|^2+
\frac12\,\log ({\rm e}^{-\phi_{10}}\,s_2)+\frac12\,
\log ( U_2^{(3)} T_2^{(3)}\, |\ell_{a}^{(3)}|^2) \right)
~.
\end{equation}
Now the logarithmic term in (\ref{s1loginst3}) correctly accounts for field
theory zero-mode contribution and the remaining finite term is the ``primed''
part of the annulus contribution that appears in the instantonic amplitudes,
namely
\begin{equation}
\label{11}
\mathcal{A}_{5_a}'\,=\,- 8\pi^2k \,{\widetilde \Delta}_a~.
\end{equation}

To discuss the holomorphic properties of the $k$-instanton induced effective
action we have to first rewrite the above expression in terms of the
supergravity variables (\ref{stu}), getting
\begin{equation}
\label{del2}
\begin{aligned}
\mathcal{A}_{5_a}' \,=\,&-{b_1\,k}\left(\ii\,\varphi+\log |\eta(u^{(3)})|^2+
\frac12\,\log (s_2)+
\log ( u_2^{(3)} t_2^{(3)}\, |\ell_{a}^{(3)}|^2) \right)\\
=\,&-(2N_a-N_F)\,k\left(\ii\,\varphi+\log |\eta(u^{(3)})|^2\,-\,\frac12\, \log
(g_a^2)\,-\,\frac12\,\log K_{\Phi} \right)~,
\end{aligned}
\end{equation}
where in the second line we have made use of Eqs. (\ref{gyma}) and
(\ref{zc}). Thus, the part of the prefactor in the instanton
amplitudes that comes from the annulus diagrams is
\begin{equation}
\label{enro}
{\rm e}^{\mathcal{A}_{5_a}'}\,=\,\left(
|\eta(u^{(3)})|^2\,\ee^{\ii\,\varphi}\right)^{-(2N_a-N_F)k} \,
(g_a\,\sqrt{K_{\Phi}})^{(2N_a-N_F)k}~.
\end{equation}
This is one of the main results in this paper. It shows that the non holomorphic
terms produced by the instanton annulus amplitudes nicely combine in the
K\"ahler metric of the adjoint fields (see also Ref. \cite{Akerblom:2007uc}) and
precisely cancel the prefactor $(g_a\,\sqrt{K_{\Phi}})^{(N_F-2N_a)k}$ in the
non-perturbative effective action (\ref{sk2}) which is produced by the rescaling
from the string basis to the supergravity basis.

Furthermore, by tuning the (arbitrary) phase $\varphi$ of the IR cutoff to be
$\arg\big(\eta(u^{(3)})^2\big)$, we can promote the harmonic term
$|\eta(u^{(3)})|^2$ to a purely holomorphic one $\eta(u^{(3)})^2$. Thus, the
$k$-instanton induced effective action (\ref{sk2}) acquires its final form
\begin{eqnarray}
S_k = {\Lambda'}^{(2N_a-N_F)k}\,
\Bigg\{\int d^4x_0 \,d^2\theta&&\Big[\frac1{2g_a^2}\,\tau_{uv}(\Phi,M)
\,W^\alpha_u
W_{\alpha v}\Big]
\nonumber\\
+&&\int d^4x_0\, d^2\theta\,d^2\bar\theta
\,\Big[K_\Phi\,\bar \Phi_u \Phi_u^{\mathrm{D}}(\Phi,M)\Big]\Bigg\}~,
\label{sk3}
\end{eqnarray}
where we have performed the rescaling
\begin{equation}
\Lambda'\,=\, \Lambda\, \eta (u^{(3)} )^{-2}
\label{lam'}
\end{equation}
which is equivalent to the following holomorphic redefinition of the
Wilsonian Yang-Mills coupling constant:
\begin{equation}
\tau_{\mathrm{YM}}\equiv\left(\frac{\theta_{\mathrm{YM}}}{2\pi}+
\ii\,\frac{4\pi^2}{g_a^2}\right)\,\rightarrow
\,\tau_{\mathrm{YM}}+\ii\,\frac{(2N_a-N_F)}{2\pi}\,\log(\eta(u^{(3)}))^2~.
\label{snew}
\end{equation}
The effective action (\ref{sk3}) has the holomorphic structure required by
supersymmetry in Wilsonian actions \cite{Kaplunovsky:1994fg,Louis:1996ya}. This
result is also a confirmation of the K\"ahler metrics (\ref{zc1}) and
(\ref{kQ1}) for the adjoint and flavored fields.

\section{Conclusions}
\label{sec:concl}
The detailed analysis of the previous sections shows that the instantonic
annulus amplitudes have the right structure to reproduce the appropriate
K\"ahler metric dependence in such a way that the instanton induced effective
action becomes purely holomorphic in the variables of the supergravity basis. To
further elaborate on this point, it is instructive to consider separately the
charged and flavored 1-loop amplitudes ${\mathcal A}_{5_a;9_a}$ and ${\mathcal
A}_{5_a;9_b}$, given respectively in Eqs. (\ref{atotfin}) and (\ref{atotfinba}),
and rewrite them in terms of the K\"ahler metrics $K_\Phi$ and $K_Q$ of the
adjoint and fundamental chiral multiplets. Using (\ref{zc1}) and (\ref{kQ1}), as well
as the coupling constant (\ref{gyma}) and the bulk K\"ahler potential
(\ref{kpot}), we easily find
\begin{eqnarray}
\label{a5a9ak}
\mathcal{A}_{5_a;9_a}
&=& - N_a\,k
\left(\log \frac{\mu^2}{M_P^2}\,+\,
\log\big(\eta(u^{(3)})\big)^4\,-\, \log (g_a^2)\,-\,\log K_{\Phi} \right)~,
\\
\label{a5a9bk}
\mathcal{A}_{5_a;9_b}
&=& \frac{N_F\,k}{2}\,
\left(\log \frac{\mu^2}{M_P^2}\,+\,
\log \big(\eta(u^{(3)})\big)^4\,-\, K + 2\log K_Q \right)~,
\end{eqnarray}
where the phase of the complex IR cutoff has been chosen as
discussed in the previous section. These two formulas are
particular cases of the expression of the one-loop running coupling constant $g^2(\mu)$
given in \cite{Kaplunovsky:1994fg,Louis:1996ya,Shifman:1986zi}. This expression can be written
in terms of the corresponding one-loop amplitude $\mathcal{A}$, according to the discussion in section
\ref{subsec:margc}%
\footnote{One has to use the fact that that $1/g^2=-\mathrm{Re}(\mathcal{A})/8\pi^2k$.},
as follows%
\begin{equation}
\mathcal{A}
= k\left[-\frac{b}2\,\log \frac{\mu^2}{M_P^2}\,+\,f\,+\,\frac{c}{2}\,K
\,-\,T(G)\,\log\left(\frac{1}{g^2}\right)+\sum_r n_r\,T(r)\,\log
K_r\right]~,
\label{kl}
\end{equation}
where $f$ is a holomorphic quantity, $K$ is the bulk K\"ahler potential and
\begin{equation}
\begin{aligned}
&T(r)\,\delta_{AB}=\Tr_r\big(T_AT_B\big)~,\quad
T(G)=T(\mathrm{adj})~,
\\
&b=3\,T(G)-\sum_r n_r\, T(r)~,\quad c=T(G)-\sum_r n_r\,T(r)~,
\end{aligned}
\end{equation}
with $T^A$ being the generators of the gauge group $G$ and $n_r$ the number of
$\mathcal{N}=1$ chiral multiplets in representation $r$, having K\"ahler metric $K_r$.
In fact, the charged
annulus amplitude (\ref{a5a9ak}) corresponds to the case of the adjoint matter
($b=2N_a$, $c=0$) while the flavored amplitude (\ref{a5a9bk}) corresponds to
$2N_F$ chiral multiplets in the fundamental representation ($b=c=-N_F$). In both
cases, $f$ is proportional to $\log(\eta(u^{(3)})^2$ and represents a finite
holomorphic renormalization of the Wilsonian Yang-Mills coupling.

Even if we have considered models with $\mathcal{N}=2$
supersymmetry, throughout this paper we have mostly used a
$\mathcal{N}=1$ notation, and also Eqs. (\ref{a5a9ak}) -
(\ref{kl}) have been written in this language. However, it is not
difficult to convert them to a full-fledged $\mathcal{N}=2$
notation. To this aim, let us observe that in (\ref{a5a9bk})
the dependence on $t_2^{(1)}$, $t_2^{(2)}$, $u_2^{(1)}$ and $u_2^{(2)}$
actually drops out, so that we can express the result in terms of the
$\mathcal{N}=2$ bulk K\"ahler potential \cite{Berg:2005ja}
\begin{equation}
{\widetilde K} = K - 2\log K_Q = -\log (s_2) - \log(t_2^{(3)}) -
\log(u_2^{(3)})
\label{kpot2}
\end{equation}
without introducing a K\"ahler metric for the hyper-multiplets. In
this way we see that both Eqs. (\ref{a5a9ak}) and (\ref{a5a9bk})
are two particular cases of the formula
\cite{Kaplunovsky:1994fg,Louis:1996ya,Shifman:1986zi,de Wit:1995zg}
\begin{equation}
\mathcal{A}
= k\left[-\frac{b}2\,\log \frac{\mu^2}{M_P^2}\,+\,f
\,-\,T(G)\,\log\left(\frac{1}{g^2}\right)
\,+\,T(G)\,\log\left(K_\Phi\right)
-\sum_r N_r\,T(r)\,{\widetilde K}\right]~,
\label{kl2}
\end{equation}
where $b$ is again the coefficient of the $\beta$-function and
$N_r$ is the number of $\mathcal{N}=2$ hyper-multiplets in the
representation $r$. Notice also that in terms of the K\"ahler potential (\ref{kpot2}),
Eq. (\ref{rel2}) can be written as
\begin{equation}
\ee^{\widetilde{K}/2}= g_a\sqrt{K_\Phi}~,
\label{rel22}
\end{equation}
while Eq. (\ref{kl2}) becomes
\begin{equation}
\mathcal{A}
= k\left[f+\frac{b}{2} \Big(\log \frac{M_P^2}{\mu^2} + \widetilde K \Big)
\right]~,
\end{equation}
which are in the appropriate form required by $\mathcal{N}=2$
supergravity \cite{de Wit:1995zg}.

We conclude by stressing that
the general formula (\ref{kl}) allows to obtain the explicit expression of the
K\"ahler metrics $K_r$ starting from an instantonic annulus amplitude
$\mathcal{A}$ in a gauge theory with a specified matter content. This can be
particularly useful in the case of $\mathcal{N}=1$ models in which the K\"ahler
metric of flavored chiral multiplets is not known {\it a priori} since they
correspond to string excitations of twisted sectors. Applying the formula
(\ref{kl}) to $\mathcal{N}=1$ theories and using it to check the holomorphicity
of the non-perturbative superpotential terms induced by instantons will
therefore provide a way to deduce the K\"ahler metric for the twisted matter in
$\mathcal{N}=1$ theories. This will be the subject of a separate publication
\cite{n1}.

\acknowledgments{We thank C. Bachas, M. Bianchi, B. de Wit, F. Fucito,
D. Luest, F. Morales, R. Russo and  A. Tanzini for many useful discussions.
This work is partially supported by the Italian MUR under
contract PRIN-2005023102 {``Strings, D-branes and Gauge Theories''} and by the
European Commission FP6 Programme under contract MRTN-CT-2004-005104
``{Constituents, Fundamental Forces and Symmetries of the Universe}'', in which
A.L. is associated to University of Torino, and R.M to INFN-Frascati. We thank
the Galileo Galilei Institute for Theoretical Physics for the hospitality and
the INFN for partial support during the completion of this work.}

\appendix
\section{Calculation of the integral $I$}
\label{integral}
In this appendix we give some details on the explicit calculation of the
integral
\begin{equation}
\label{integr-r0l}
 I \equiv
\int_0^{\infty} \frac{d\tau}{\tau}
\sum_{(r_1,\,r_2)  \in\mathbb{Z}^2}
{\rm e}^{ -\,2 \pi \tau
\frac{| r_1 U^{(3)}\,- \,r_2 |^2 }{U_2^{(3)}T_2^{(3)} |\ell_a^{(3)}|^2}}~.
\end{equation}

\paragraph{Regularization with a cut-off}
In the IR region ($\tau \rightarrow \infty$) the integral (\ref{integr-r0l})
has a logarithmic divergence due to the massless states
and a regularization procedure is necessary to cure the IR problem.
Here we use the regularization procedure introduced in Ref. \cite{Dixon:1990pc}
and insert in the integrand the regulator
\begin{equation}
\label{a1}
R(\tau)= 1 - \ee^{-\frac{\pi }{\alpha'\,m^2\,\tau}}~,
\end{equation}
where  $m$ is a (complex) IR cut-off. In the following we will briefly discuss
another regularization scheme with Wilson lines.

Eq. (\ref{integr-r0l}) is divergent also in the UV-region $\tau\rightarrow 0$.
This divergence was not present in Ref. \cite{Dixon:1990pc} and
in order to cure it  we use the Poisson resummation formula to rewrite
Eq. (\ref{integr-r0l}) in the form:
\begin{equation}
\label{a2}
I \equiv
\frac{|\ell_a^{(3)}|^2 \,T_2^{(3)}}{2}\!
\int_0^{\infty}\! \frac{d \tau}{ \tau^2}\!\sum_{(s_1,\,s_2)
\in\mathbb{Z}^2-\{(0,0)\} } \!\!\!\!
{\rm e}^{-\frac{\pi}{2\tau}\,
\frac{ |\ell_a^{(3)}|^2 T^{(3)}}{U_2^{(3)}}
| U^{(3)} s_1 \,+\,\,s_2 |^2 }
\left(1 - \ee^{-\frac{\pi }{\alpha'\,m^2\,\tau}}\right)~,
\end{equation}
where we have neglected  the divergent contribution due to the term
$s_1=s_2=0$,  because it is absent in a consistent model free of tadpoles
\cite{Lust:2003ky}.

We can now perform the integral getting:
\begin{equation}
\label{ku}
I =
\frac{U_2^{(3)}}{\pi} \sum_{(s_1,\,s_2) \in\mathbb{Z}^2-\{(0,0)\} }
\left[ \frac{1}{|U^{(3)} s_1 + s_2 |^2} -
\frac{1}{|U^{(3)} s_1 + s_2 |^2 +\,U_2^{(3)}\,N }\right]
\end{equation}
with $N= 2/(\alpha'm^2\,T_2^{(3)}|\ell_a^{(3)}|^2)$.
By using the identity:
\begin{equation}
\begin{aligned}
\sum_{s_2 \in \mathbb{Z} }
\frac{ 1 }{ (s_2 + A )^2 + B^2 }
&=
\frac{\ii\, \pi}{2 B} \left[ \cot \pi(A+\ii\,B) - \cot\pi (A-\ii\,B) \right]
\\
&=-\frac{\pi}{B}
\left[
\frac{ \ee^{2 \pi \ii\,u} }{\ee^{2\pi\ii\, u} -1}
+
\frac{ \ee^{-2 \pi\ii\, \bar u} }{ \ee^{-2 \pi\ii\, \bar u} -1}
-1\right]
\simeq\,\frac{\pi}{B}
\quad\mbox{for}~B\rightarrow +\infty
\end{aligned}
\label{fund-seriesl}
\end{equation}
with $u=A+\ii\,B$, $A=U_1^{(3)}s_1$ and
$B=U_2^{(3)}s_1$ (or $B=\sqrt{(U_2^{(3)}s_1)^2+ NU_2^{(3)}})$, we have:
\begin{eqnarray}
I &=& -2\sum_{s_1>0}
\left[
\frac{1}{s_1}\frac{ q^{s_1} }{ q^{s_1} -1 }
+\frac{1}{s_1}\frac{ \bar q^{s_1} }{ \bar q^{s_1} -1 }
\right]
+
 \sum_{s_1>0}
\left[
\frac{2}{s_1}- \frac{2}{\sqrt{s_1^2 +\frac{N}{U_2}}}
\right]
\nonumber\\
&&
+\frac{U_2^{(3)}}{\pi}
\sum_{s_2\in \mathbb{Z}-\{0\}}
\left[
\frac{1}{ s_2^2}
-\frac{1}{ s_2^2 +N U_2^{(3)} }
\right]
\end{eqnarray}
with $q=\ee^{2\pi\ii\, U^{(3)}}$.
Expanding the geometric series, the first term gives
\begin{equation}
 -2\sum_{s_1>0}
\frac{1}{s_1}\frac{ q^{s_1} }{ q^{s_1} -1 }
\,=\, 2\sum_{n,s_1>0} \frac{1}{s_1} q^{n s_1}
-\log ( q^{-1/6} \eta(U^{(3)})^2)~,
\end{equation}
where $\eta(U)$ is the Dedekind  $\eta$-function. The second term can be
evaluated by using the Euler-Maclaurin formula:
\begin{equation}
 2\sum_{s_1>0}
\left[
\frac{1}{s_1}- \frac{1}{\sqrt{s_1^2 +\frac{N}{U_2^{(3)}}}}
\right] \, \simeq \, 2 \log \frac{\sqrt{N}}{2 \sqrt{U_2^{(3)}} } + 2 \gamma_E~.
\end{equation}
The last term yields:
\begin{equation}
\frac{U_2^{(3)}}{\pi}
\sum_{s_2\in \mathbb{Z}-\{0\}}
\left[
\frac{1}{ s_2^2}
-\frac{1}{ s_2^2 +N }
\right]
\,=\,2\frac{U_2^{(3)}}{\pi}\zeta(2) - O(m^2)\simeq + \frac{\pi}{3} U_2^{(3)}~,
\end{equation}
where we used the particular value of Riemann zeta function $\zeta(2)=\pi^2/6$.
Finally we can write:
\begin{equation}
I \,=\, - \log  |\eta(U^{(3)}) |^4
- \log\left( U_2^{(3)}  T_2^{(3)} |\ell_a^{(3)}|^2  \right)
- \log \left(\alpha' m^2\right)
\label{ku1}
\end{equation}
where we have redefined $2m^2\ee^{-2\gamma_E} \to m^2$.

\paragraph{Regularization with Wilson lines}
We now briefly describe the effect of introducing Wilson lines on the torus
$\mathcal{T}_2^{(3)}$ which can act as IR regulators \cite{Abel:2006yk} for the
integral $I$ in Eq. (\ref{integr-r0l}).

Turning on Wilson lines $\xi_1$ and $\xi_2$ along $\mathcal{T}_2^{(3)}$ produces
a shift on the momenta so that $I$ becomes
\begin{equation}
\label{kint+wilson}
K(\xi_1,\xi_2) \equiv
\int_0^{\infty} \frac{d\tau}{\tau}
\sum_{(r_1,\,r_2)  \in\mathbb{Z}^2}
{\rm e}^{ -\,2 \pi \tau
\frac{| (r_1-\xi_1) U^{(3)}\,-\,
(r_2-\xi_2) |^2 }{U_2^{(3)}T_2^{(3)} |\ell_a^{(3)}|^2}}~.
\end{equation}
Subtracting the UV divergence after a Poisson resummation as we did
before in Eq. (\ref{a2}), we have
\begin{equation}
\label{a2w}
K(\xi_1,\xi_2) \equiv
\frac{|\ell_a^{(3)}|^2 T_2^{(3)}}{2}\!
\int_0^{\infty}\! \frac{d \tau}{ \tau^2}\!\sum_{(s_1,\,s_2)
\in\mathbb{Z}^2-\{(0,0)\} } \!\!\!\!
{\rm e}^{-\frac{\pi}{2\tau}\,
\frac{ |\ell_a^{(3)}|^2 T^{(3)}}{U_2^{(3)}}
| U^{(3)} s_1 \,+\,\,s_2 |^2 \,+\,2\pi\ii(s_1\xi_1+s_2\xi_2)}
\end{equation}
which can be easily integrated to give
\begin{equation}
K(\xi_1,\xi_2)= \frac{U_2^{(3)}}{\pi}
\sum_{(s_1,\,s_2)
\in\mathbb{Z}^2-\{(0,0)\} }
\frac{ \ee^{2 \pi \ii (s_1 \xi_1 + s_2 \xi_2)}  }{|U^{(3)} s_1 + s_2
|^2}~.
\label{W-ku}
\end{equation}
If $\xi_2=0$ we can use Eq. (\ref{fund-seriesl}) and write
\begin{eqnarray}
K(\xi_1,\xi_2=0) &=&
 -\sum_{s_1>0}
\left[\frac 1{s_1}\,
\frac{ q^{s_1} ( \ee^{2\pi\ii \,\xi_1 s_1}+ \ee^{-2\pi\ii \,\xi_1 s_1})}{q^{s_1}
-1}
+
\frac 1{s_1}\,\frac{ \bar q^{s_1} (\ee^{2\pi\ii\, \xi_1 s_1}+
\ee^{-2\pi \ii\,\xi_1 s_1})}{\bar q^{s_1} -1 }
\right]
\nonumber\\
&&+
 \sum_{s_1>0}
\frac{ ( \ee^{2\pi\ii\, \xi_1 s_1}+ \ee^{- 2\pi\ii\, \xi_1 s_1}) }{s_1}
+\frac{U_2^{(3)}}{\pi}
\sum_{s_2\in \mathbb{Z}-\{0\}}
\frac{1}{ s_2^2}
\label{a45}
\end{eqnarray}
with $q= \ee^{2\pi\ii U^{(3)}}$. Expanding the geometric series, the first term
gives
\begin{equation}
-\sum_{s_1>0}
\frac{1}{s_1}
\frac{ q^{s_1} ( \ee^{2\pi\ii\, \xi_1 s_1}+ \ee^{-2\pi\ii\, \xi_1 s_1})
}{ q^{s_1} -1 }
=-\sum_{n>0} \log\left((1-  \ee^{2\pi\ii\, \xi_1}\,q^{n })
                     (1-  \ee^{- 2\pi\ii \xi_1}\,q^{n })\right)~,
\end{equation}
and similarly for the second term with $q$ replaced by $\bar q$.
The second line of (\ref{a45}) can be easily seen to give
\begin{equation}
-\log\big(4\sin^2(\pi\xi_1)\big) + \frac{\pi}3\,U_2^{(3)}~,
\label{a46}
\end{equation}
so that we can finally write
\begin{eqnarray}
K(\xi_1,\xi_2=0)&=&
\frac{\pi}{3}\, U_2^{(3)}
- \log\Big| 2 \sin (\pi \xi_1)
\prod_{n=1}^\infty
(1- e^{2\pi\ii\,\xi_1}\, q^{n }) (1- q^{n } \ee^{- 2\pi\ii\,\xi_1})
\Big|^2
\nonumber\\
&=&
- \log \left|\frac{\theta_1\big(\xi_1| -\ii\, U^{(3)}\big)}{\eta(U^{(3)}) }
\right|^2~.
\label{W-ku-final-special}
\end{eqnarray}
To find the general expression for $\xi_2\not=0$, it is convenient
to introduce the complex variable $z=\xi_1-U^{(3)}\xi_2$ such that
$z\simeq z+1 \simeq z-U^{(3)}$ as a consequence of the periodicity
of the Wilson lines. Then, one can show that
\begin{equation}
\frac{\partial}{\partial \bar z} \frac{\partial}{\partial z}\,
K(\xi_1,\xi_2) =\frac{\pi}{U_2^{(3)}}
\Big[1 -\delta(\xi_1)~\delta(\xi_2)\Big]~.
\end{equation}
Studying the behavior of the solution to this differential
equation near $z=0$ and matching with the form (\ref{W-ku-final-special})
of the explicit solution already found for $\xi_2=0$, one can
obtain \cite{Abel:2006yk,Blumenhagen:2006ci}
\begin{equation}
K(\xi_1,\xi_2)=
- \log \left|\ee^{-\ii\pi\xi_2U^{(3)}}\,\frac{\theta_1\big(z| -\ii\,
U^{(3)}\big)}{\eta(U^{(3)}) }
\right|^2~.
\label{W-ku-final}
\end{equation}
This final result can be entirely written as the sum of a holomorphic and an anti-holomorphic function, in agreement with the fact that in the Wilson line regularization
all excitations are massive.


\begin{thebibliography}{99}

\bibitem{Polchinski:1994fq}
  J.~Polchinski,
  Phys.\ Rev.\ D {\bf 50} (1994) 6041
  [arXiv:hep-th/9407031].

\bibitem{Becker:1995kb}
  K.~Becker, M.~Becker and A.~Strominger,
  Nucl.\ Phys.\  B {\bf 456} (1995) 130
  [arXiv:hep-th/9507158].

\bibitem{Green:1997di}
  M.~B.~Green and P.~Vanhove,
  Phys.\ Lett.\  B {\bf 408} (1997) 122
  [arXiv:hep-th/9704145].

\bibitem{Green:1997as}
  M.~B.~Green, M.~Gutperle and P.~Vanhove,
  Phys.\ Lett.\  B {\bf 409} (1997) 177
  [arXiv:hep-th/9706175].

\bibitem{Kiritsis:1997em}
  E.~Kiritsis and B.~Pioline,
  Nucl.\ Phys.\  B {\bf 508} (1997) 509
  [arXiv:hep-th/9707018].

\bibitem{Bachas:1997mc}
  C.~Bachas, C.~Fabre, E.~Kiritsis, N.~A.~Obers and P.~Vanhove,
  Nucl.\ Phys.\  B {\bf 509} (1998) 33
  [arXiv:hep-th/9707126].

\bibitem{Green:1997tv}
  M.B.~Green and M.~Gutperle,
  Nucl.Phys. B {\bf 498} (1997) 195
  [arXiv:hep-th/9701093];
  JHEP {\bf 9801} (1998) 005 [arXiv:hep-th/9711107];
  Phys. Rev. D {\bf 58} (1998) 046007
  [arXiv:hep-th/9804123].

\bibitem{Green:2000ke}
  M.B.~Green and M.~Gutperle,
  JHEP {\bf 0002} (2000) 014 [arXiv:hep-th/0002011].

\bibitem{Banks:1998nr}
  T.~Banks and M.~B.~Green,
  JHEP {\bf 9805} (1998) 002
  [arXiv:hep-th/9804170].

\bibitem{Chu:1998in}
  C.~S.~Chu, P.~M.~Ho and Y.~Y.~Wu,
  Nucl.\ Phys.\  B {\bf 541}, 179 (1999)
  [arXiv:hep-th/9806103].

\bibitem{Kogan:1998re}
  I.~I.~Kogan and G.~Luzon,
  Nucl.\ Phys.\  B {\bf 539} (1999) 121
  [arXiv:hep-th/9806197].

\bibitem{Bianchi:1998nk}
  M.~Bianchi, M.~B.~Green, S.~Kovacs and G.~Rossi,
  JHEP {\bf 9808} (1998) 013
  [arXiv:hep-th/9807033].

\bibitem{Dorey:1998xe}
  N.~Dorey, V.~V.~Khoze, M.~P.~Mattis and S.~Vandoren,
  Phys.\ Lett.\  B {\bf 442} (1998) 145
  [arXiv:hep-th/9808157];
  N.~Dorey, T.~J.~Hollowood, V.~V.~Khoze, M.~P.~Mattis and S.~Vandoren,
  JHEP {\bf 9906} (1999) 023
  [arXiv:hep-th/9810243];
  Nucl.\ Phys.\  B {\bf 552} (1999) 88
  [arXiv:hep-th/9901128].

\bibitem{Witten:1995im}
  E.~Witten,
  Nucl.\ Phys.\ B {\bf 460} (1996) 335
  [arXiv:hep-th/9510135].

\bibitem{Douglas:1996uz}
  M.~R.~Douglas,
  J.\ Geom.\ Phys.\  {\bf 28} (1998) 255
  [arXiv:hep-th/9604198].
  arXiv:hep-th/9512077.

\bibitem{Dorey:2002ik}
  N.~Dorey, T.~J.~Hollowood, V.~V.~Khoze and M.~P.~Mattis,
  Phys.\ Rept.\  {\bf 371} (2002) 231
  [arXiv:hep-th/0206063].

\bibitem{Bianchi:2007ft}
  M.~Bianchi, S.~Kovacs and G.~Rossi,
  arXiv:hep-th/0703142.

\bibitem{Billo:2002hm}
  M.~Billo, M.~Frau, I.~Pesando, F.~Fucito, A.~Lerda and A.~Liccardo,
  JHEP {\bf 0302} (2003) 045
  [arXiv:hep-th/0211250].

\bibitem{Di Vecchia:1997pr}
  P.~Di Vecchia, M.~Frau, I.~Pesando, S.~Sciuto, A.~Lerda and R.~Russo,
  Nucl.\ Phys.\  B {\bf 507} (1997) 259
  [arXiv:hep-th/9707068].

\bibitem{Billo:2004zq}
  M.~Billo, M.~Frau, I.~Pesando and A.~Lerda,
  JHEP {\bf 0405} (2004) 023
  [arXiv:hep-th/0402160].

\bibitem{Billo:2005fg}
  M.~Billo, M.~Frau, S.~Sciuto, G.~Vallone and A.~Lerda,
  JHEP {\bf 0605} (2006) 069
  [arXiv:hep-th/0511036].

\bibitem{Billo:2006jm}
  M.~Billo, M.~Frau, F.~Fucito and A.~Lerda,
  JHEP {\bf 0611} (2006) 012
  [arXiv:hep-th/0606013].

\bibitem{Beasley:2005iu}
  C.~Beasley and E.~Witten,
  JHEP {\bf 0602} (2006) 060
  [arXiv:hep-th/0512039].

\bibitem{Blumenhagen:2006xt}
  R.~Blumenhagen, M.~Cvetic and T.~Weigand,
  Nucl.\ Phys.\  B {\bf 771} (2007) 113
  [arXiv:hep-th/0609191].

\bibitem{Ibanez:2006da}
  L.~E.~Ibanez and A.~M.~Uranga,
  JHEP {\bf 0703} (2007) 052
  [arXiv:hep-th/0609213].

\bibitem{Abel:2006yk}
  S.~A.~Abel and M.~D.~Goodsell,
  arXiv:hep-th/0612110.

\bibitem{Akerblom:2006hx}
  N.~Akerblom, R.~Blumenhagen, D.~L{\"{u}}st, E.~Plauschinn and M.~Schmidt-Sommerfeld,
  JHEP {\bf 0704} (2007) 076
  [arXiv:hep-th/0612132].

\bibitem{Bianchi:2007fx}
  M.~Bianchi and E.~Kiritsis,
  arXiv:hep-th/0702015.

\bibitem{Cvetic:2007ku}
  M.~Cvetic, R.~Richter and T.~Weigand,
  arXiv:hep-th/0703028.

\bibitem{Argurio:2007vq}
  R.~Argurio, M.~Bertolini, G.~Ferretti, A.~Lerda and C.~Petersson,
  JHEP {\bf 0706} (2007) 067
  [arXiv:0704.0262 [hep-th]].

\bibitem{Bianchi:2007wy}
  M.~Bianchi, F.~Fucito and J.~F.~Morales,
  arXiv:0704.0784 [hep-th].

\bibitem{Ibanez:2007rs}
  L.~E.~Ibanez, A.~N.~Schellekens and A.~M.~Uranga,
  arXiv:0704.1079 [hep-th].

\bibitem{Akerblom:2007uc}
  N.~Akerblom, R.~Blumenhagen, D.~Lust and M.~Schmidt-Sommerfeld,
  arXiv:0705.2366 [hep-th].

\bibitem{Antusch:2007jd}
  S.~Antusch, L.~E.~Ibanez and T.~Macri,
  arXiv:0706.2132 [hep-ph].

\bibitem{Blumenhagen:2007zk}
  R.~Blumenhagen, M.~Cvetic, D.~Lust, R.~Richter and T.~Weigand,
  arXiv:0707.1871 [hep-th].

\bibitem{Aharony:2007pr}
  O.~Aharony and S.~Kachru,
  arXiv:0707.3126 [hep-th].

\bibitem{Blumenhagen:2007bn}
  R.~Blumenhagen, M.~Cvetic, R.~Richter and T.~Weigand,
  arXiv:0708.0403 [hep-th].

\bibitem{Blumenhagen:2006ci}
  R.~Blumenhagen, B.~Kors, D.~Lust and S.~Stieberger,
  arXiv:hep-th/0610327.

\bibitem{Kaplunovsky:1994fg}
  V.~Kaplunovsky and J.~Louis,
  Nucl.\ Phys.\  B {\bf 422} (1994) 57
  [arXiv:hep-th/9402005].
  Nucl.\ Phys.\  B {\bf 444} (1995) 191
  [arXiv:hep-th/9502077].

\bibitem{Louis:1996ya}
  J.~Louis and K.~Foerger,
  Nucl.\ Phys.\ Proc.\ Suppl.\  {\bf 55B} (1997) 33
  [arXiv:hep-th/9611184].

\bibitem{Shifman:1986zi}
  M.~A.~Shifman and A.~I.~Vainshtein,
  Nucl.\ Phys.\  B {\bf 277} (1986) 456.

\bibitem{Derendinger:1991kr}
  J.~P.~Derendinger, S.~Ferrara, C.~Kounnas and F.~Zwirner,
  Phys.\ Lett.\  B {\bf 271} (1991) 307;
  Nucl.\ Phys.\  B {\bf 372} (1992) 145.

\bibitem{Lopes Cardoso:1992yd}
  G.~Lopes Cardoso and B.~A.~Ovrut,
  Nucl.\ Phys.\  B {\bf 392} (1993) 315
  [arXiv:hep-th/9205009].

\bibitem{Ibanez:1992hc}
  L.~E.~Ibanez and D.~Lust,
  Nucl.\ Phys.\  B {\bf 382} (1992) 305
  [arXiv:hep-th/9202046].

\bibitem{Gaillard:1992bt}
  M.~K.~Gaillard and T.~R.~Taylor,
  Nucl.\ Phys.\  B {\bf 381} (1992) 577
  [arXiv:hep-th/9202059].

\bibitem{Binetruy:1991sz}
  P.~Binetruy, G.~Girardi and R.~Grimm,
  Phys.\ Lett.\  B {\bf 265}, 111 (1991).

\bibitem{de Wit:1995zg}
  B.~de Wit, V.~Kaplunovsky, J.~Louis and D.~Lust,
  Nucl.\ Phys.\  B {\bf 451} (1995) 53
  [arXiv:hep-th/9504006].

\bibitem{Hollowood:2002ds}
  T.~J.~Hollowood,
  JHEP {\bf 0203} (2002) 038
  [arXiv:hep-th/0201075].

\bibitem{Lust:2003ky}
  D.~L{\"{u}}st and S.~Stieberger,
  arXiv:hep-th/0302221.

\bibitem{Douglas:1996du}
  M.~R.~Douglas and M.~Li,
  arXiv:hep-th/9604041.

\bibitem{Bachas:1996zt}
  C.~Bachas and C.~Fabre,
  Nucl.\ Phys.\  B {\bf 476} (1996) 418
  [arXiv:hep-th/9605028].

\bibitem{Di Vecchia:2005vm}
  P.~Di Vecchia, A.~Liccardo, R.~Marotta and F.~Pezzella,
  Int.\ J.\ Mod.\ Phys.\  A {\bf 20} (2005) 4699
  [arXiv:hep-th/0503156].

\bibitem{Lust:2004cx}
  D.~L{\"{u}}st, P.~Mayr, R.~Richter and S.~Stieberger,
  Nucl.\ Phys.\  B {\bf 696} (2004) 205
  [arXiv:hep-th/0404134].

\bibitem{Antoniadis:1996vw}
  I.~Antoniadis, C.~Bachas, C.~Fabre, H.~Partouche and T.~R.~Taylor,
  Nucl.\ Phys.\  B {\bf 489} (1997) 160
  [arXiv:hep-th/9608012].

\bibitem{Berg:2005ja}
  M.~Berg, M.~Haack and B.~K\"ors,
  JHEP {\bf 0511} (2005) 030
  [arXiv:hep-th/0508043].

\bibitem{Bertolini:2005qh}
  M.~Bertolini, M.~Billo, A.~Lerda, J.~F.~Morales and R.~Russo,
  Nucl.\ Phys.\  B {\bf 743} (2006) 1
  [arXiv:hep-th/0512067].

\bibitem{Nekrasov:2002qd}
  N.~A.~Nekrasov,
  Adv.\ Theor.\ Math.\ Phys.\  {\bf 7} (2004) 831
  [arXiv:hep-th/0206161];
  R.~Flume and R.~Poghossian,
  Int.\ J.\ Mod.\ Phys.\ A {\bf 18} (2003) 2541
  [arXiv:hep-th/0208176];
  R.~Flume, F.~Fucito, J.~F.~Morales and R.~Poghossian,
  JHEP {\bf 0404} (2004) 008
  [arXiv:hep-th/0403057].

\bibitem{Di Vecchia:2003ae}
  P.~Di Vecchia, A.~Liccardo, R.~Marotta and F.~Pezzella,
  JHEP {\bf 0306} (2003) 007
  [arXiv:hep-th/0305061].

\bibitem{D'Adda:1977ur}
  A.~D'Adda and P.~Di Vecchia,
  Phys.\ Lett.\  B {\bf 73} (1978) 162.

\bibitem{Billo:1998vr}
  M.~Billo, P.~Di Vecchia, M.~Frau, A.~Lerda, I.~Pesando, R.~Russo and
  S.~Sciuto,
  Nucl.\ Phys.\  B {\bf 526} (1998) 199
  [arXiv:hep-th/9802088].

\bibitem{Billo:2000yb}
  M.~Billo, B.~Craps and F.~Roose,
  JHEP {\bf 0101} (2001) 038
  [arXiv:hep-th/0011060].

\bibitem{n1}
  M.~Billo, P.~Di Vecchia, M.~Frau, A.~Lerda, R.~Marotta and I.~Pesando,
  arXiv:0709.0245 [hep-th].

\bibitem{Dixon:1990pc}
L.~J.~Dixon, V.~Kaplunovsky and J.~Louis,
  Nucl.\ Phys.\  B {\bf 355} (1991) 649.

\end{thebibliography}
\end{document}